\documentclass[fleqn,usenatbib]{mnras} 

\usepackage{amsmath}
\usepackage{array}

\usepackage{graphicx}	
\usepackage{amsmath}	
\usepackage{amssymb}	
\usepackage{pdflscape}
\usepackage[section]{placeins}
\usepackage{epstopdf}

\newcommand{\Msun}{\mathrm{M_{\odot}}}
\newcommand{\Rsun}{\mathrm{R_{\odot}}}
\newcommand{\Lsun}{\mathrm{L_{\odot}}}

\title[CHIRON investigation of long-period binaries]{Orbital and physical parameters of eclipsing binaries from the ASAS catalogue -- XI. CHIRON investigation of long-period binaries\thanks{Based in part on data collected through CNTAC proposals CN-2012B-036, CN-2013A-093, CN-2013B-022, CN-2014A-044, CN-2014B-067, CN-2015A-074; ESO programmes 078.D-0245, 087.C-0012, 088.D-0080, 089.C-0415, 089.D-0097, 090.D-0061, 091.D-0145, 094.C-0428; as well as TESS Guest Investigator programme G011083.}}

\author[M. Ratajczak et al.]{M. Ratajczak,$^{1,2}$\thanks{E-mail: milena@ncac.torun.pl} R.~K. Paw{\l}aszek,$^{3}$ K.~G. He{\l}miniak,$^{3}$ M. Konacki,$^{3}$
\newauthor P. Sybilski,$^{3}$ S.~K. Koz{\l}owski,$^{3}$ M. Litwicki,$^{3} $A.~M.~S. Smith,$^{4}$ \newauthor P.~Miko{\l}ajczyk,$^{2}$ D.~R.~Anderson,$^{5}$ and C. Hellier${^5}$\\
$^{1}$Warsaw University Observatory, Al. Ujazdowskie 4, 00-478 Warszawa, Poland\\
$^{2}$Astronomical Institute, University of Wroc{\l}aw, Kopernika 11, 51-622 Wroc{\l}aw, Poland\\
$^{3}$Nicolaus Copernicus Astronomical Center, Polish Academy of Sciences, ul. Rabia\'{n}ska 8, 87-100 Toru\'{n}, Poland\\
$^{4}$Institute of Planetary Research, German Aerospace Center, Rutherfordstrasse 2, 12489 Berlin, Germany\\
$^{5}$Astrophysics Group, Lennard-Jones Laboratories, Keele University, Keele, Staffordshire, ST5 5BG, UK}

\date{Accepted XXX. Received YYY; in original form ZZZ}

\pubyear{2019}

\begin{document}
\label{firstpage}
\pagerange{\pageref{firstpage}--\pageref{lastpage}}
\maketitle
  
\begin{abstract}
We present the results of a spectroscopic campaign on eclipsing binaries with long orbital period ($P = 20-75$ d) carried out with the CHIRON spectrograph. Physical and orbital solutions for seven systems were derived from the \textit{V}-band, and \textit{I}-band ASAS, WASP, and TESS photometry, while radial velocities were calculated from high quality optical spectra using a two-dimensional cross-correlation technique. The atmospheric parameters of the stars have been determined from the separated spectra.
Most of our targets are composed of evolved stars (sub-giants or red giants) but two systems show components in different phases of evolution and one possible merger. For four binaries the masses and radii of the components were obtained with precision better than $3\%$.
These objects provide very valuable information on stellar evolution.
\end{abstract}




\section{Introduction}

Long-period binary systems ($P > 8$ d) are useful tools to study several aspects of the theory of stellar evolution. Since they may consist of detached red giant (RG) stars, issues like convection or stellar activity may be investigated. In order to reject inadequate models \citep{tor10} with sufficiently strong constraints, observational data must yield parameters (like stellar masses and radii) with errors lower than $3\%$. With precise measurements of radial velocity (RV) and reliable light curves (LC), detached eclipsing binaries (DEBs) can deliver such quality.

Of 244 systems from the catalogue of physical properties of well-studied eclipsing binaries DEBCat \citep{sou15}, about thirty pairs consist of sub-giant or RG components \citep[e.g. ][]{and91, hel15, suc15, bro18, gra18, suc19}. As late-type DEBs are amongst the best candidates for distance determinations \citep{pie13}, the vast majority of well-characterised evolved systems are located in the Magellanic Clouds. Only for some of them the atmospheric parameters were obtained out of spectral reconstruction techniques. The most valuable systems providing stringent tests of theoretical stellar evolutionary models are those with components in different phases of evolution. Literature quotes only a few, including the extensively studied AI Phe \citep{tor10, max20} and TZ For \citep{hig18}. Evolved systems can be also used to test the absolute calibration of the surface brightness-colour relation for late-type stars \citep{gal16}.

In addition to eclipsing binaries, RGs are also studied in non-eclipsing systems, like $\alpha$ Aur \citep{tor15} with highly precise ($0.3\%$) masses determination.
Furthermore, astroseismology has opened a new way of characterising pulsating RGs in eccentric binary systems \citep[e.g. ][]{bec14} and validate already determined stellar parameters with independent methods \citep{fra13}.

This paper is part of a larger effort to describe a rich sample of eclipsing binaries from the \textit{All Sky Automatic Survey} \citep*[ASAS,][]{poj97} catalogue \citep{hel09,hel11a,rat13,hel14,hel15,rat16,hel19}. We hereby present the results of a spectroscopic campaign on long-period binaries carried out with the CHIRON spectrograph \citep{sch12,tok13} in order to identify and characterise more sub-giants and RGs in DEBs. The results include the first detailed studies (with spectral analysis of separated spectra) of the DEBs A-051753, V643 Ori, A-061016, A-062926, A-065114, A-090232, A-110814, and the discussion of the evolutionary stages of their components. We shall first describe the targets, then the data collection and analysis, and finally show the results we obtained.

\section{Targets}

All of our targets were chosen from the \textit{ASAS Catalog of Variable Stars} \citep*[ACVS,][]{poj02} using the late-type binaries selection criteria described in our previous papers \citep{rat13,rat16}: orbital period $P > 8$ d (to include detached components of solar radius or larger), amplitude of brightness changes in $V$ lower than $1$ mag (to exclude classical Algol systems), and $V-K > 1$ mag (to focus on late-type stars). The selection comprises seven systems, whose physical and orbital characteristics were determined out of spectroscopy, providing an estimate of the evolutionary status and age too.

The analysed sample include the following systems:

\begin{itemize}
    \item A-051753 (CPD-54~810, ASAS~J051753-5406.0)
    \item V643 Ori (HD~294651, ASAS~J060700.9-025458)
    \item A-061016 (CD-33~2771, ASAS~J061016-3321.3)
    \item A-062926 (TYC 6511-1799-1, ASAS~J062926-2513.5)
    \item A-065114 (HD~265111, ASAS~J065114+0753.9)
    \item A-090232 (TYC~8590-374-1, ASAS~J090232-5653.4)
    \item A-110814 (TYC~8620-1809-1, ASAS~J110814-5555.4)
\end{itemize}

Their apparent ACVS $V$-band magnitudes, amplitudes of photometric variations in $V$-band, colour indices $V-K$, orbital periods $P$ and \textit{Gaia} Data Release 2 \citep[GDR2;][]{gai16,gai18} effective temperatures are given in Table~\ref{tab_targets_info}. We emphasize that GDR2 temperatures are derived for single objects, not binary components.

Although there is no detailed analysis in literature for the majority of the systems, V643~Ori and A-061016 have been studied before. Apart from ACVS, V643~Ori was also listed in the following catalogues of binary stars: \citet{bra80}, \citet{avv13} and \citet{pau15}. A preliminary analysis of the system is presented in \citet{imb87}, where it is described as a K2III + K7III pair with masses of components of $3.3~\Msun$ and $1.9~\Msun$, and period $P = 54.2$ d. Due to relatively low precision of the RV measurements in the early analysis, we discarded those and only used our new measurements here. V643~Ori is also mentioned in \citet{egg16} as a potential former triple system. A-061016 was inspected by \citet{par09} looking for chromospherically active binaries, but no traces of activity were detected. It was found that the target is composed of stars of spectral types K5III and K5V. The system was also analysed as part of a survey on abundances by \citet{luc15}, who estimated the atmospheric parameters of the primary component: $T_\mathrm{eff}= 4002$ K, $\log g= 2.5$ cm~s$^{-2}$, $v_\mathrm{micro} = 1.68$ km~s$^{-1}$ and [M/H] = 0.25, however these values were derived assuming only the primary spectrum was detectable.
Moreover, literature points out that A-065114 is an X-ray source listed in the ROSAT All Sky Survey, displaying coronal activity \citep{szc08,kir12}.

\begin{table}
\begin{center}
\caption{\textit{V}-band magnitudes, amplitudes of photometric variations $\Delta V$, colour indices $V-K$, orbital periods $P$ from ACVS and GDR2 effective temperatures for the analysed systems.}
\label{tab_targets_info}
\begin{tabular}{c>{\centering\arraybackslash}m{0.8cm}>{\centering\arraybackslash}m{0.8cm}>{\centering\arraybackslash}m{0.8cm}cc}
\hline
\hline
Object ID & $V$ [mag] & $\Delta V$ [mag] & $V{-}K$ [mag] & $P$ [d] & $T_\mathrm{GDR2}$ [K] \\

\hline
\rule{0pt}{3ex} A-051753 & $10.45$ & $0.14$ & $1.17$ &  $26.132$ & $6\,495\substack{+187 \\ -141}$ \\
\rule{0pt}{3ex} V643 Ori &  $9.35$ & $0.78$ & $3.03$ &  $52.48$  & $4\,434\substack{+101 \\  -72}$ \\
\rule{0pt}{3ex} A-061016 &  $9.71$ & $0.46$ & $3.64$ & $199 $    & $3\,998\substack{+329 \\  -72}$ \\
\rule{0pt}{3ex} A-062926 & $11.43$ & $0.24$ & $1.36$ &  $26.382$ & $6\,215\substack{+408 \\ -154}$ \\
\rule{0pt}{3ex} A-065114 &  $9.22$ & $0.24$ & $2.59$ &  $43.5$   & $4\,632\substack{+174 \\ -130}$ \\
\rule{0pt}{3ex} A-090232 & $10.63$ & $0.02$ & $1.12$ &  $20.822$ & $6\,555\substack{+240 \\ -117}$ \\
\rule{0pt}{3ex} A-110814 & $10.62$ & $0.24$ & $3.0$  &  $75.228$ & $4\,507\substack{+108 \\ -134}$ \\
\hline
\end{tabular}
\end{center}
\end{table}

\section{Observational data}

\subsection{Photometry}

For the preliminary LC analysis we used the ASAS \textit{V}-band photometry. A total number of $1\,317$, $518$, $867$, $483$, $254$, $633$, and $521$ measurements were available in the ACVS for A-051753, V643 Ori, A-061016, A-062926, A-065114, A-090232, A-110814, respectively. Additionally we used ASAS \textit{I}-band photometry \citep{sit14} for A-061016, which contained $658$ data points. 

A-051753, A-061016, and A-062926 were also observed by the Wide Angle Search for Planets (WASP) southern instrument \citealt{pol06}. WASP-South is located at the South African Astronomical Observatory (SAAO), and consists of eight Canon $200$~mm $f/1.8$ lenses, each equipped with a broadband filter ($400 -- 700$~nm), and an Andor $2048\times2048$ e2V CCD camera, on a single robotic mount. A-051753 was observed a total of $16\,454$ times, A-061016 $23\,596$ times, and A-062926 $21\,692$ times.

Moreover, for A-051753, A-062926, and A-090232, we have used TESS \citep[Transiting Exoplanet Survey Satellite,][]{ric15} $2$-minute cadence measurements. The targets were observed $78\,042$ (sectors $3$ to $7$), $14\,828$ (sector $6$), and $29\,216$ (sectors $8$ and $9$) times, respectively. The rest of our targets was either not well covered (e.g. part of only one eclipse), or not observed by the telescope yet. TESS photometric data were obtained via the Mikulski Archive for Space Telescopes (MAST) service operated by the Space Telescope Science Institute (STScI). 

\subsection{Spectroscopy}

Most of our spectroscopic data was obtained with the CHIRON spectrograph. CHIRON is a fibre-fed Echelle spectrometer at the $1.5$~m telescope operated by the Small and Moderate Aperture Research Telescope System Consortium \citep[SMARTS;][]{sub10} in Cerro Tololo, Chile. A spectral resolution of $R \sim 80\,000$ is achieved with an image slicer, however a slit mask with $R \sim 90\,000$ and $136\,000$ (that incurs in larger light losses) is also available. For our purposes we mostly used the fibre mode which gives $R \sim 25\,000$. We aimed at a signal-to-noise ratio of $\sim 50$ at $\lambda=5\,500$~\AA.

Additional spectra were obtained with the $3.6$-m ESO telescope equipped with the High Accuracy Radial velocity Planet Searcher (HARPS) spectrograph \citep[$R \sim 115\,000$;][]{may03}, the $2.2$-m MPG telescope with the Fiberfed Extended Range Optical Spectrograph \citep[FEROS;][]{kau99} reaching $R\sim 48\,000$, the $1.2$-m Euler telescope with the CORALIE spectrograph \citep[$R \sim 60\,000$;][]{que01}, and the $3.9$-m AAT telescope equipped with the University College London Echelle Spectrograph \citep[UCLES; ][]{die90} with $R\sim 60\,000$. Altogether we analysed $23$, $29$, $33$, $20$, $19$, $11$, and $21$ spectra for A-051753, V643 Ori, A-061016, A-062926, A-065114, A-090232, A-110814, respectively.

\section{Data analysis}

A detailed data analysis procedure is described in the previous papers from the series: \citet{rat13,rat16}. It includes data reduction, RVs measurements, and a modelling method. The data used in this study are presented in Tab.~\ref{tab_rms}: in addition to ASAS photometry available for all  targets, TESS data was collected for three systems (A-051753, A-062926, A-090232) and one has a WASP light curve (A-061016). As the photometric observations are not homogeneous (the quality and quantity of the TESS measurements are much higher than the ASAS and WASP datasets), we used the set of better quality for the final analysis.

\subsection{Data reduction}

ASAS \textit{V} and \textit{I} photometric data, reduced by the ASAS pipeline, were downloaded in form of brightness measurements with uncertainties from the following catalogues: ACVS and \textit{ASAS 3 -- The Catalogue of Bright Variable Stars in I-band South of Declination of +28$^{\circ}$}.

The WASP data were reduced by the WASP reduction pipeline \citep{pol06}. Red noise was removed with the algorithm by \citet{tam05}.

The analysis of the TESS Full Frame Images (FFIs) with $30$~min sampling time was done for all the targets having photometry available with $2$~min cadence. Comparing the two sets of data we found very high consistency but naturally used the dataset with better sampling rate. The downloaded light curves (sectors $3-7$, $6$ and $8-9$ for A-051753, A-062926, and A-090232, respectively) were initially cleaned employing some quality flags per measurement. Next the orbital frequency, along with hundreds of harmonic frequencies, was fitted to the data. Splines were used to fit the residuals and de-trend the original data. During this process a significant portion of the data for A-051753 had to be removed due to an enormous trend in sector $4$. After de-trending $57\,944$, $12\,785$ and $29\,097$ data points were available for further analysis for A-051753, A-062926 and A-090232, respectively.

As spectroscopic data were obtained using several instruments, we used various algorithms to reduce and calibrate the spectra. CHIRON data were reduced with the pipeline by \citet{tok13} and \textsc{iraf}\footnote{\textsc{iraf} http://iraf.noao.edu/ is written and supported by the \textsc{iraf} programming group at the National Optical Astronomy Observatories (NOAO) in Tucson, AZ. NOAO is operated by the Association of Universities for Research in Astronomy (AURA), Inc. under cooperative agreement with the National Science Foundation.}, while the \textit{rvsao.bcvcorr} method was used for corrections on barycentric velocity and time. For the data from FEROS and CORALIE we used the pipeline by \citet{jor14}. HARPS spectra were treated with the \textit{Data Reduction Software} (DRS), while UCLES data were reduced with the standard \textsc{iraf} procedures.

\subsection{RV measurements}
\label{rv}

In order to determine the RVs of the stars we applied our own implementation of two dimensional cross-correlation technique (TODCOR) \citep{zuc94}. Initially, we took synthetic spectra as reference, computed using \textsc{atlas9} and \textsc{atlas12} codes \citep{kur92}. However the final RVs were calculated using separated spectra as reference (lower values of RV formal errors). The obtained RV fitting RMS value was comparable to the case of using synthetic spectra.

Formal RV errors were computed by applying the bootstrap analysis \citep{pre07} of TODCOR correlation maps (1\,000 bootstrap samples). As every spectrum we analysed was divided into echelle spectrum orders, the maps were created by adding randomly selected maps from each spectral order. Formal errors were multiplied by an appropriate factor to avoid underestimation and to obtain the best-fit with reduced $\chi^2 \approx$ 1 (see Sec.~\ref{rv_fit}).

RV measurements for both components, their final errors, O-Cs, exposure times per spectrum, signal-to-noise ratio per collapsed spectral pixel at $\lambda=5\,500$~\AA, and instrument specifications are collected in Appendix A \ref{RV_table_051753}--\ref{RV_table_110814}. 

\subsection{Modelling tools}

The following procedures and codes were used for modelling the systems:
\begin{itemize} 
\item
\textsc{v2fit} \citep{kon10} -- to find the spectroscopic orbit from RV data,
\\
\item
\textsc{jktebop} \citep{sou04a,sou04b} -- to model LCs,
\\
\item
\textsc{phoebe} \citep[\textit{Physics Of Eclipsing Binaries};][]{prs05} --- to check consistency of RV and LC solutions  and estimate the preliminary effective temperature ratio,
\\
\item
\textsc{sme} \citep[\textit{Spectroscopy Made Easy};][]{val96,val98} to perform spectral analysis of the reconstructed spectra,
\\
\item
\textsc{jktabsdim} \citep{sou04a,sou04b} -- to calculate absolute values of parameters.
\\
\end{itemize}

As described in the previous papers of our series, the primary component is assumed to undergo eclipsing during the deeper (primary) eclipse, for a time $T_0$. For eccentric systems $T_0$ is set to coincide with the phase of conjunction when the argument of the periastron $\omega$ is $\pi / 2$. 

A detailed description of the modelling procedures used in the following steps of the analysis is given in the next subsections.

\subsection{RV fitting}
\label{rv_fit}

In order to find the spectroscopic solution we used our own procedure \textsc{v2fit} which fits a double-Keplerian orbit to RV data and minimises the $\chi^2$ function with the Levenberg-Marquardt algorithm. As mentioned in Sec.~\ref{rv}, RV errors were rescaled to cure underestimation and obtain a fit with reduced $\chi^2 \approx 1$.

Whenever data sets were produced by different spectrographs, we calibrated by matching the RV zero points, obtained by fitting an additional parameter. We minimised the risk of obscuring real long-term RV shifts by choosing the longest datasets as the base RV-set (CORALIE for A061016 and CHIRON for the rest).

In those cases where the centre-of-mass velocities $\gamma_{i=1,2}$ of the components exhibited non-negligible differences, we took
$\gamma_1$ as the final velocity of the system centre-of-mass by subtracting $\gamma_2$ - $\gamma_1$ from the RV of the secondary. Differences in velocity have been pointed out in previous DEBs studies \citep{tor09} and may be explained by large-scale convective motions \citep{sch75, por00} that could be different in the two stars (and are poorly characterised for giant stars) or by the presence of spots. 

In this step we obtained the system mass ratio $q$, semimajor axis $a$, RV semi-amplitudes $K_1$ and $K_2$, and the systemic velocity $\gamma$. Their uncertainties were estimated with a bootstrap analysis. Preliminary values of orbital period $P$, zero-phase time $T_0$ (see Sec.~\ref{lc_fit}), eccentricity $e$, and longitude of periastron $\omega$ were also computed at this stage. 
When the values of $\gamma_2$ - $\gamma_1$ and $e$ were not significantly far from zero, we set them to $0$ for further use.

\subsection{LC fitting}
\label{lc_fit}

In order to model LC data, we used \textsc{jktebop} (version 2.8), which utilises the Levenberg-Marquardt optimisation algorithm to find the best-fitting model. Its advantage for eccentric systems is to avoid falling in the strong correlation problem of eccentricity $e$ and longitude of periastron $\omega$ by fitting non-correlated combination terms $e \cos \omega$ and $e \sin \omega$, which depend on times of minima and duration of the eclipses. In this step the final values of the orbital period $P$, inclination $i$, eccentricity $e$, longitude of periastron $\omega$, surface brightness ratio, and fractional radii of the components were calculated. As formal errors (from the covariance matrix found by the minimisation algorithm) are underestimated, $10\,000$ iterations of a Monte Carlo algorithm were used to collect statistics on the parameters and yield the final errors.

For the LC analysis the gravitational darkening coefficients were set to $\beta = 0.32$ \citep{luc67} and the limb darkening was modelled with the logarithmic law of \citet{kli70} with coefficients taken from \citet{van93}. After the initial checking, we assumed no third light for all cases.

For systems with more than one LC available (e.g. from ASAS-\textit{V}, WASP, and TESS) the dataset of better quality was used to determine the values of LC-dependent parameters and estimate their errors.

The consistency of RV and LC solutions was checked simultaneously in various pass-bands by using \textsc{phoebe}. The preliminary effective temperature ratio was estimated too. The first estimation of effective temperature of the brighter component was based on the colour-temperature calibration \citep{wor11} using TYCHO-2 colours \citep{hog00}. However, for the final analysis we used the temperatures obtained by performing the spectral analysis with \textsc{sme}.

\subsection{Spectral separation and analysis}

The phase coverage of CHIRON spectra was sufficient to perform spectral separation. In order to obtain reconstructed spectra of both components, we used the tomographic method of \citet{bag91}, numerically formulated and solved by the code described in \citet{kon10}. In this technique, a sequence of spectra of the binary as a whole is employed to reconstruct the spectra of both stars, with their RVs given as input. No information on the orbital parameters is needed.

This spectral separation can be applied irrespective of the initial flux ratio of the stars, thus we renormalised the spectra by a factor dependent on the brightness ratio (BR). BR values were taken as the mean of the individual BRs, calculated for every spectral order of every spectrum (obtained as extra output of the TODCOR analysis). A-061016 was handled differently, as the spectrum taken during a total eclipse enabled independent estimation of the the BR value (found in agreement with that from the TODCOR analysis; see also Sec.~\ref{br}).

As line blending is more severe in the blue and in cooler stars \citep[see][]{val05}, thus making continuum placement and derived parameters less accurate, we used a spectral range of $5\,927$ to $6\,399$~\AA~and \textit{Vienna Atomic Line Database} \citep[VALD;][]{pis95,kup99} as input to the spectral analysis with \textsc{sme}. In principle the procedure fits a stellar spectrum (reconstructed in our case) with a synthetic spectrum to determine stellar parameters, like effective temperature $T_\mathrm{eff}$, metallicity [M/H], projected rotational velocity $v \sin i$, surface gravity $\log g$, and microturbulent velocity $v_\mathrm{micro}$.

\begin{figure}
\includegraphics[width=.35\textwidth, angle=-90, scale=0.97]{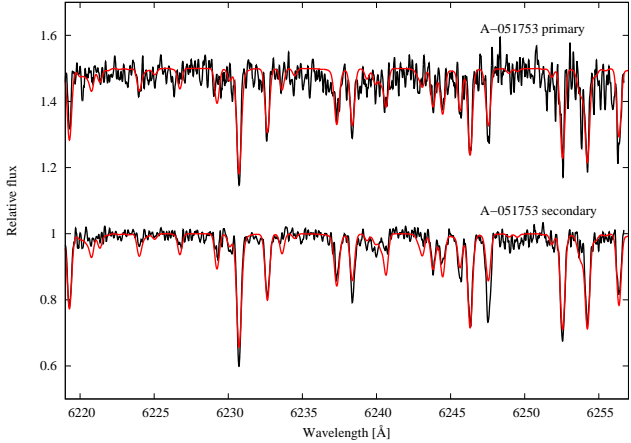}
	\caption{Separated and scaled spectra of the A-051753 system components (black) compared with synthetic spectra calculated for atmospheric parameters from spectral synthesis (red).} 
	\label{sp_comp_sme}
\end{figure}

Keeping the values of $\log g$ that we determined in the first stage of the analysis, we obtained $T_\mathrm{eff}$, [M/H] and $v \sin i$ for each of the $7$ orders in the given spectral range and averaged them to get the final results. The variance on the orders served as uncertainty. Due to partial degeneration of $v_\mathrm{micro}$ and [M/H] we decided to fix $v_\mathrm{micro} = 1.4$ km~s$^{-1}$ \citep{zie12} to minimise errors in [M/H]. Default \textsc{sme} abundance pattern taken from \citet{gre07} was also fixed. It should be noted that $v \sin i$ values lower than $\sim 10$ km~s$^{-1}$ are affected by additional error due to the insufficient resolution of the spectrograph. Examples of reconstructed separated spectra compared with the models for the estimated atmospheric parameters are presented in Fig.~\ref{sp_comp_sme}.

\subsection{Absolute stellar parameters}

The final values of the stellar parameters, together with their reliable uncertainties, were calculated by \textsc{jktabsdim} and based on the values (and errors) obtained with \textsc{v2fit} ($q$, $a$, $K$, $\gamma$), \textsc{jktebop} ($P$, $e$, $i$, fractional radii -- from the LC of best quality) and \textsc{sme} ($T_\mathrm{eff}$, [M/H]). With this code we calculated masses $M_{1}$, $M_{2}$, radii $R_{1}$, $R_{2}$, surface gravities $\log g_{1}$, $\log g_{2}$, luminosities $\log L_{1}$, $\log L_{2}$, bolometric magnitudes $M_\mathrm{bol1}$, $M_\mathrm{bol2}$ (with uncertainties), as well as distances. It should be noted that the first calculation was done without $T_\mathrm{eff}$ and yielded values of $\log g$, later used in \textsc{sme}. The second run of \textsc{jktabsdim}, with temperatures included, supplied the final values of $L$, $M_\mathrm{bol}$, and distance. The final orbital and physical parameters of the stars together with their robust error estimates are shown in Table~\ref{tab_orb}.

\subsection{Evolutionary stage and age estimation}

The evolutionary status of the stars was checked using Yonsei-Yale \citep[hereafter YY;][]{yi01} evolutionary tracks for scaled-solar mixture (no enhancement of $\alpha$ process elements). We initially assumed both stars have the same metallicity (systemic [M/H]) and estimated its value either as the average of metallicities ([M/H]$_1$ and [M/H]$_2$) or as one of them (see Sec. \ref{met}). The age of each system was obtained by fitting our data with isochrones, generated from three models: YY, PARSEC \citep[PAdova \& TRieste Stellar Evolution Code;][]{bre12}, and Dartmouth \citep{dot07}. The evolutionary stage (HR diagram) and age estimates (three planes: $M_\mathrm{bol}$, $T_\mathrm{eff}$, $\log g$ -- $M$) are given for each individual target in Fig.~\ref{051753_plots} - \ref{110814_plots} for A-051753, V643 Ori, A-061016, A-062926, A-065114, A-090232, and A-110814, respectively.

Under the assumption of systemic metallicity, the age was estimated from the isochrone best fitting the data and its uncertainty propagated from that on metallicity. Taking the minimum and maximum values of metallicity compatible with error bars, two more isochrones were fitted, giving the minimal and maximal ages of the system. The error on age was thus computed as their difference.

\section{Results}

In this Section we present the results of the modelling we obtained for A-051753, V643 Ori, A-061016, A-062926, A-065114, A-090232, and A-110814 systems. Table~\ref{tab_rms} shows the mean formal errors of photometric measurements, and the RMS of the orbital fitting for our systems, while the orbital and physical parameters of the stars are shown in Table~\ref{tab_orb} and Table~\ref{tab_par}. Values of effective temperatures $T_\mathrm{eff}$, metallicities [M/H] and rotational velocities $v \sin i$ are the results of the spectral analysis of separated spectra. The values of $v_\mathrm{rot}$ for non-eccentric systems are in agreement with the values calculated under the assumption of tidal locking. 
The errors of the parameters in this section, as well as in Table~\ref{tab_orb} and Table~\ref{tab_par} are calculated with the \textsc{jktabsdim} procedure using the parameters and uncertainties from the RV (\textsc{v2fit}) and LC (\textsc{jktebop}) analysis described above. The errors of the parameters calculated via spectral analysis were taken from their variances over spectral orders. 

We also present LCs (for all available datasets, although the final solution is based on the best quality LC), RV curve, evolutionary tracks, and isochrones in Fig.~\ref{051753_plots} -- Fig.~\ref{110814_plots}.

\subsection{Systems with red giants components}

\subsubsection{A-061016}

\begin{figure}
	\begin{center}
	\begin{tabular}{rc}
	\includegraphics[width=.27\textwidth, angle=-90]{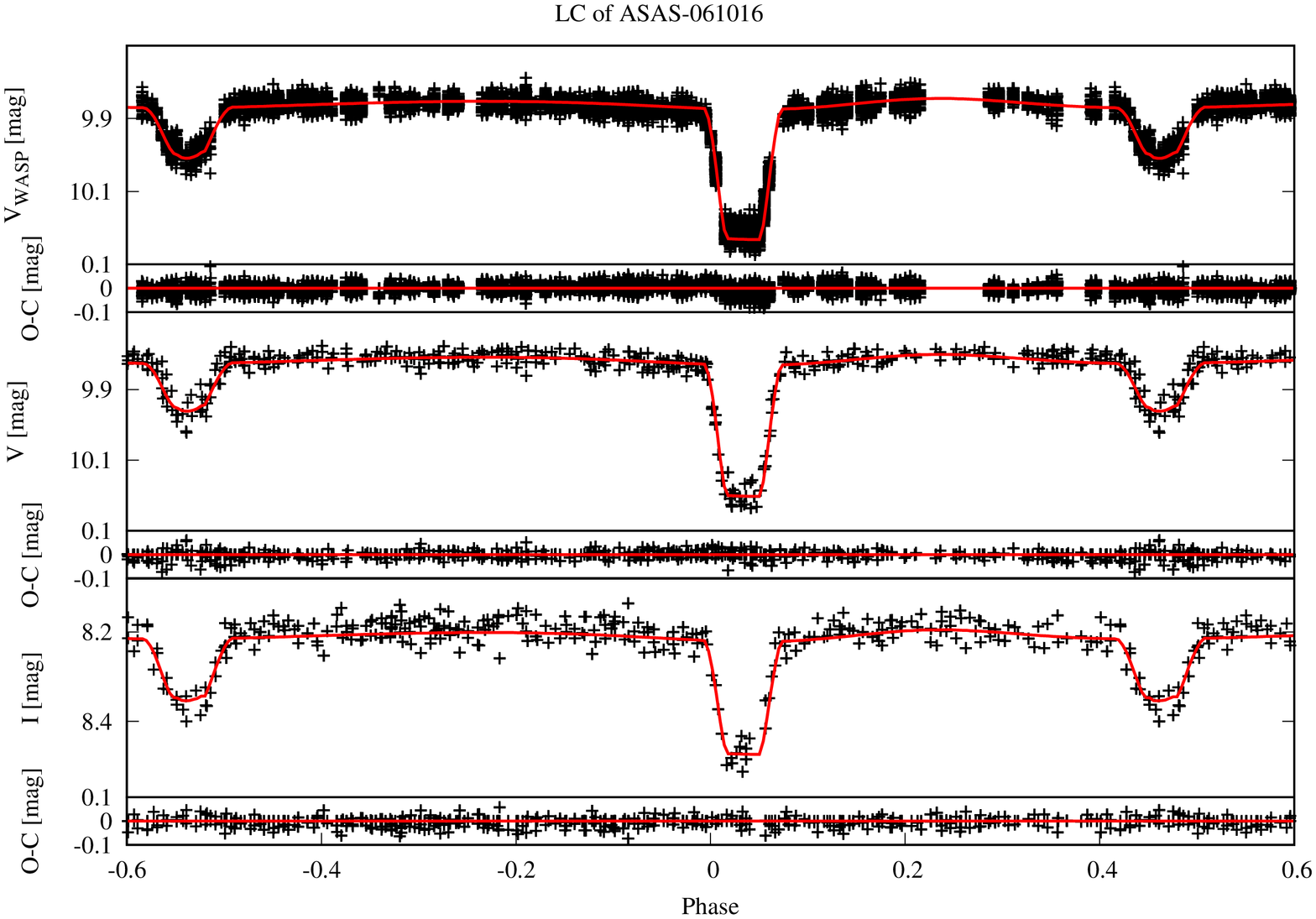}&
	\includegraphics[width=.27\textwidth, angle=-90]{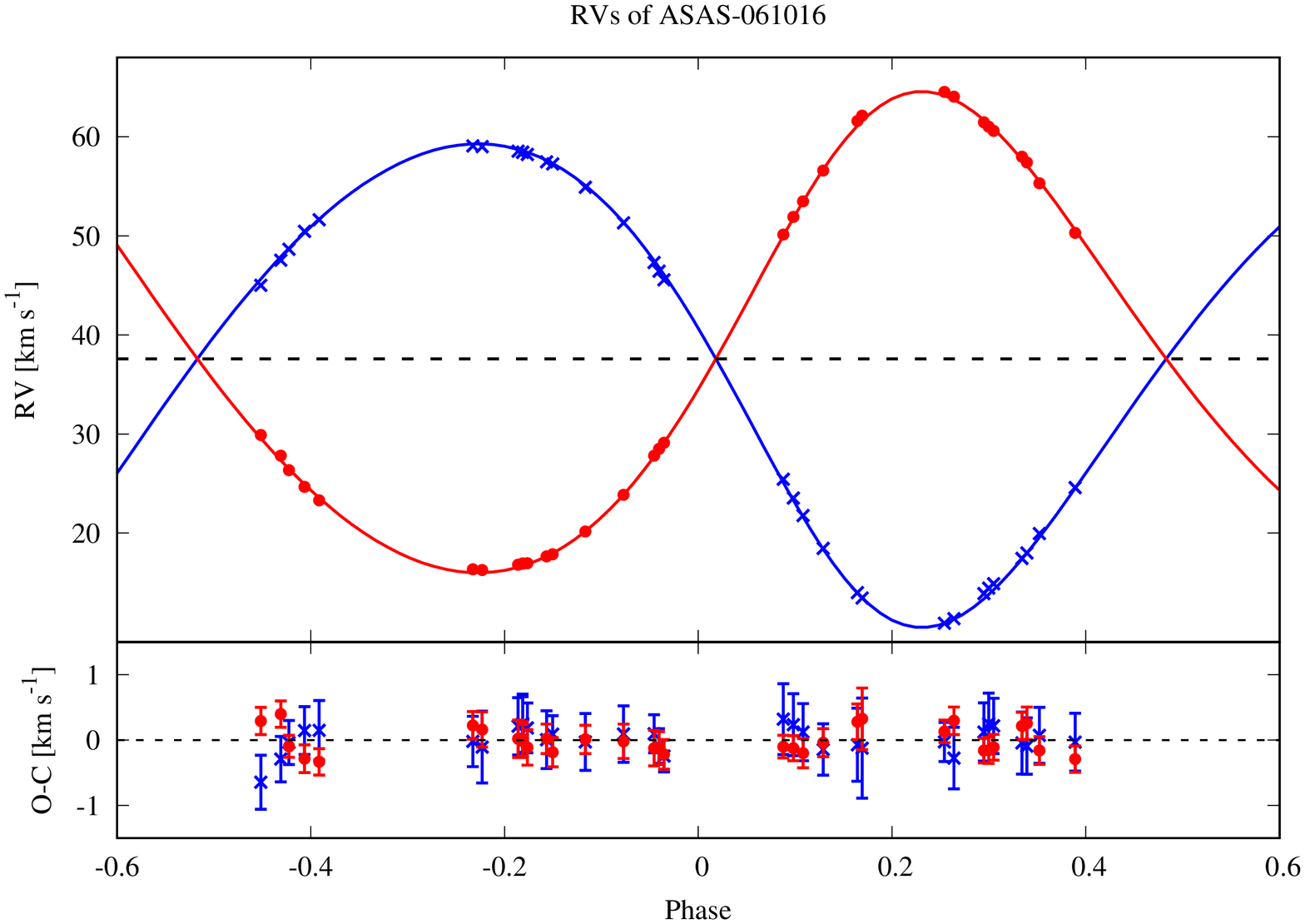}\\
	\includegraphics[width=.27\textwidth, angle=-90]{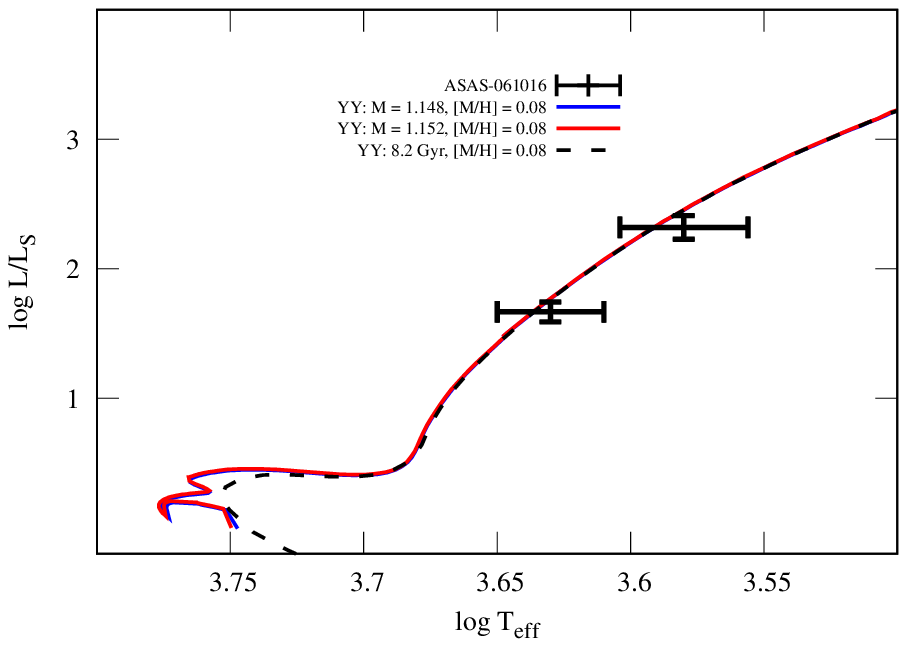}&
	\includegraphics[width=.27\textwidth, angle=-90]{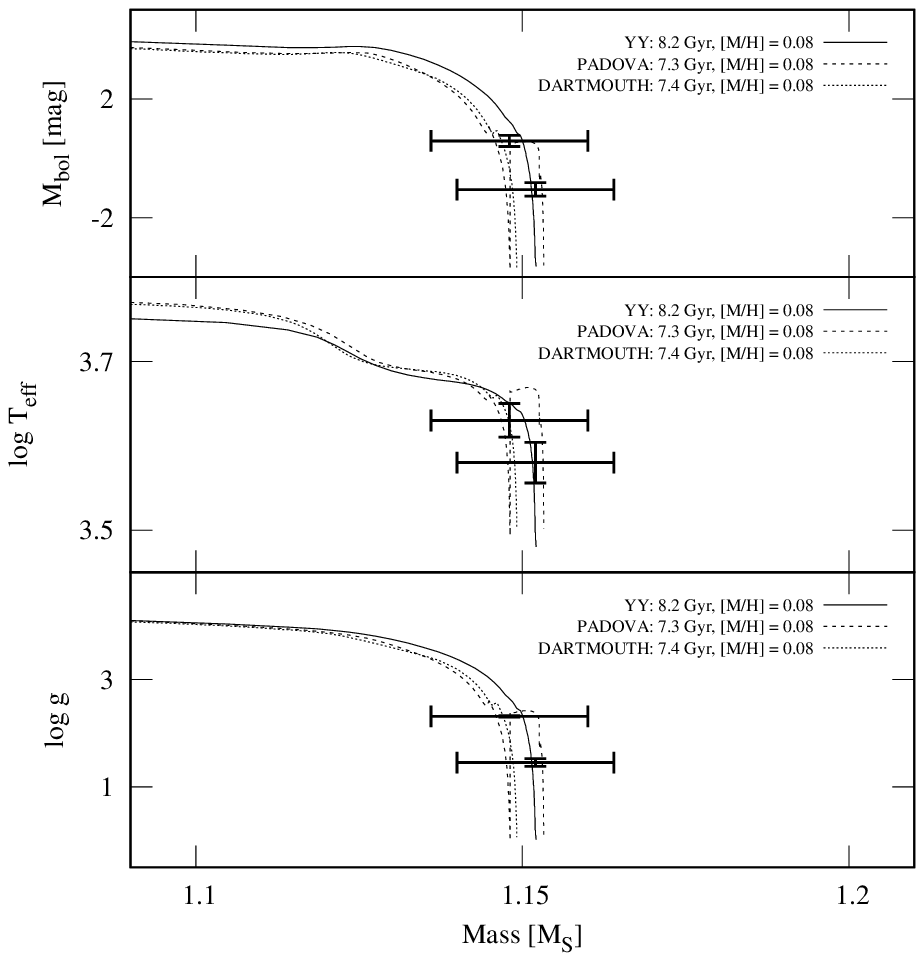}\\
	\end{tabular}
	\caption{LC (WASP, ASAS-\textit{V}, ASAS-\textit{I}; left upper panel), RV curve (upper right), evolutionary tracks (left lower) and isochrones (lower right) for  A-061016. Blue colour represents primary component, red -- secondary.} 
	\label{061016_plots}
	\end{center}
\end{figure}

The masses and radii for the components of A-061016 (see Fig.~\ref{061016_plots}) were found as $M_1 = 1.148 \pm 0.011~\Msun$, $M_2 = 1.152 \pm 0.012~\Msun$, $R_1 = 12.34 \pm 0.19~\Rsun$, and $R_2 = 33.32 \pm 0.39~\Rsun$. Further orbital and physical details are presented in Tables~\ref{tab_orb} and~\ref{tab_par}.

One of the CHIRON spectra was taken during a total eclipse (the apparent disc of the secondary fully covered the smaller primary; see also Sec.~\ref{br}), enabling a spectral analysis of the secondary with no need for spectral reconstruction. Spectral separation was instead applied to the spectra from CORALIE and FEROS.

The final values of effective temperature and metallicity were calculated as a weighted mean of the values obtained from spectral analysis of both sets of separated spectra and the spectrum taken during the total eclipse: $T_\mathrm{eff1} = 4290 \pm 190$ K, $T_\mathrm{eff2} = 3803 \pm 186$ K, [M/H] = $0.08 \pm 0.1$ . The given values of $T_\mathrm{eff}$ are in agreement with ASAS \textit{V} and \textit{I} and WASP light curve solutions (see Fig. \ref{061016_plots}), as well as with the value estimated by \citet{luc15} $T_\mathrm{eff} = 4002$ K for the primary, while the metallicity we obtained is slightly lower than [M/H]$ = 0.25$ from literature. One possible reason for this slight inconsistency is the fact that values in literature are all extracted under the assumption that only the spectrum of the primary is detectable.

Both components of A-061016 are in advanced phase of evolution (after the MS). The age of the system was estimated to be in the range $7.3-7.7 \substack{+0.45 \\ -0.35}$ Gyr (PARSEC and YY, respectively).

LC, RV curve, evolutionary tracks and isochrones can be found in Fig.~\ref{061016_plots}.

\subsubsection{A-065114}

\begin{figure}
	\begin{center}
	\begin{tabular}{cc}
\includegraphics[width=.27\textwidth, angle=-90]{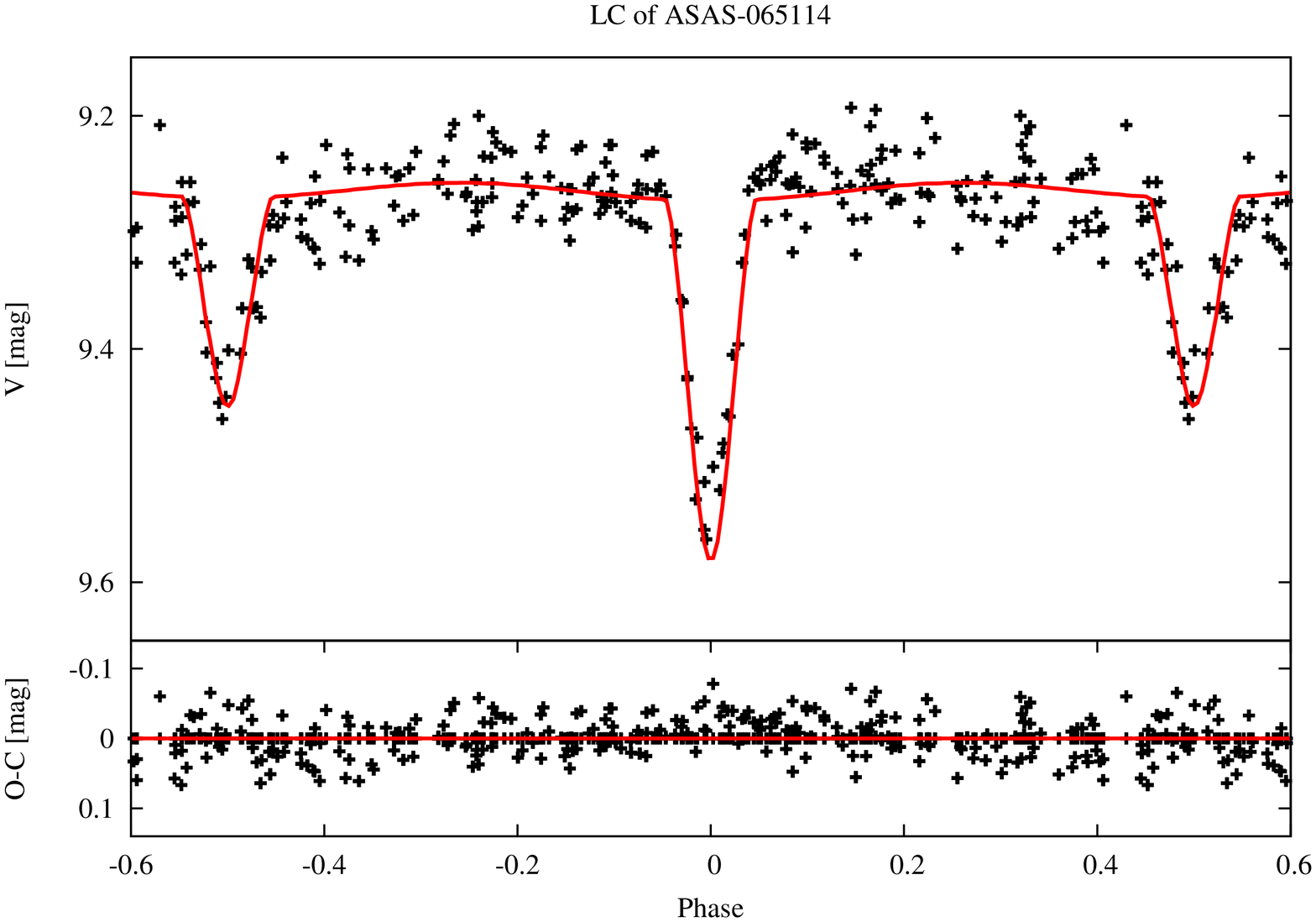}&
\includegraphics[width=.27\textwidth, angle=-90]{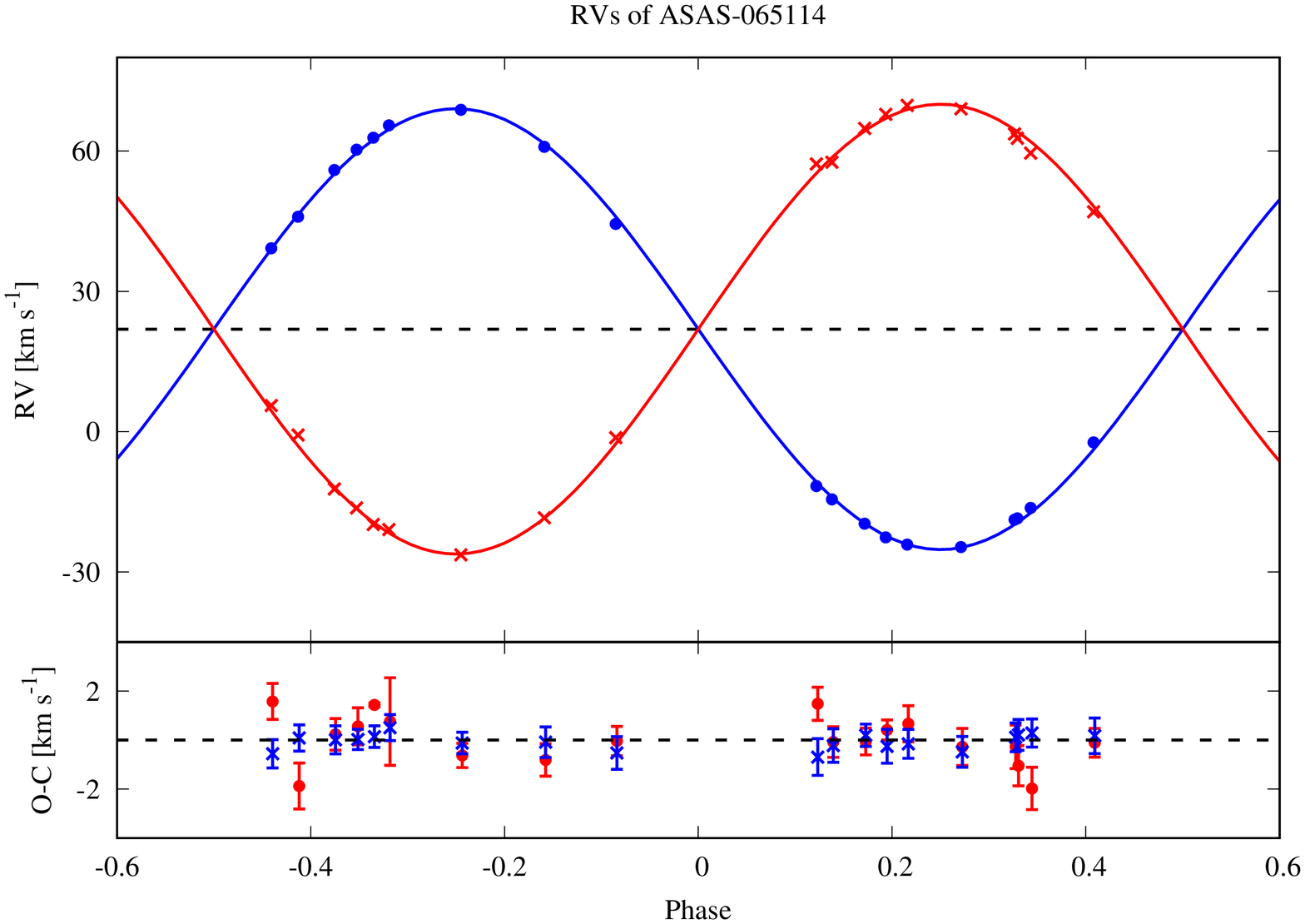}\\
\includegraphics[width=.27\textwidth, angle=-90]{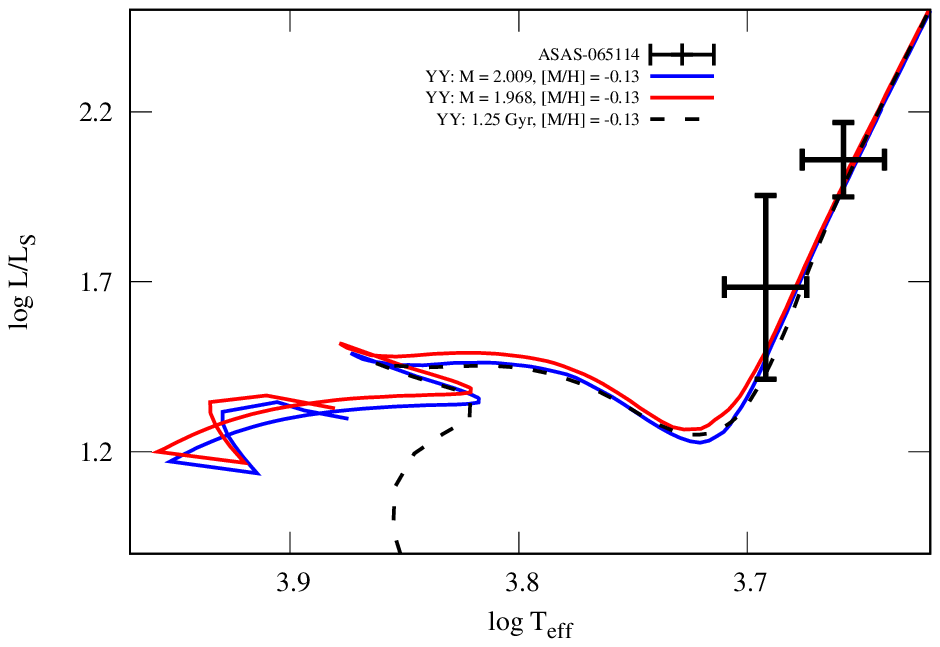}&
\includegraphics[width=.27\textwidth, angle=-90]{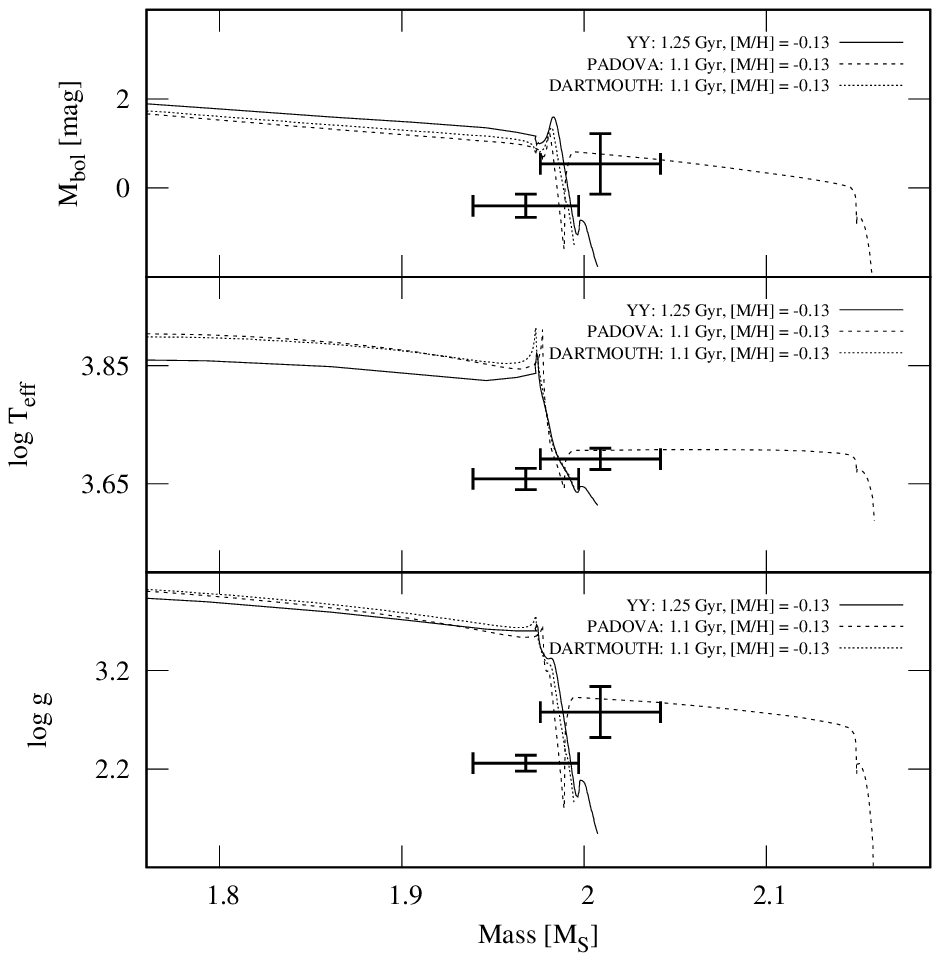}\\
	\end{tabular}
	\caption{LC (ASAS-\textit{V}; left upper panel), RV curve (upper right), evolutionary tracks (left lower) and isochrones (lower right) for A-065114. Blue colour represents primary component, red -- secondary. Green track represents an evolutionary track for $1.95 ~\Msun$ star and [M/H] = -0.1 calculated using PARSEC models.} 
	\label{065114_plots}
	\end{center}
\end{figure}

The masses and radii of the components of A-065114 (see Fig.~\ref{065114_plots}) were found as: $M_1 = 2.009 \pm 0.033~\Msun$, $M_2 = 1.968 \pm 0.029~\Msun$, $R_1 = 9.56 \pm 2.80~\Rsun$, and $R_2 = 17.23 \pm 1.48~\Rsun$. Tables~\ref{tab_orb} and~\ref{tab_par} contain the remaining physical and orbital information.

Ten CHIRON (fibre mode) spectra were separated and spectroscopically analysed. We obtained $T_\mathrm{eff1} = 4920 \pm 203$ K, $T_\mathrm{eff2} = 4550 \pm 190$ K, [M/H]$_1 = -0.16 \pm 0.15$, and [M/H]$_2 = -0.10 \pm 0.10$. Metallicities were averaged to get the final value of [M/H]$ = -0.13 \pm 0.13$.

Evolutionary tracks indicate that both stars are in advanced stage of evolution. PARSEC isochrones include the scenario where the more massive star already underwent a helium flash. Since the photometry for this system is the poorest in our sample, the errors on the radii are relatively large, and thus allow any of the mentioned evolutionary solutions. 

The age of the system was estimated to be $1.1-1.25 \substack{+0.05 \\ -0.08}$ Gyr (YY and PARSEC, respectively).

ASAS \textit{V}-band LC, RV curve, evolutionary tracks and isochrones for the system components are shown in Fig.~\ref{065114_plots}.

\subsubsection{A-110814} 

\begin{figure}
	\begin{center}
	\begin{tabular}{cc}
	\includegraphics[width=.27\textwidth, angle=-90]{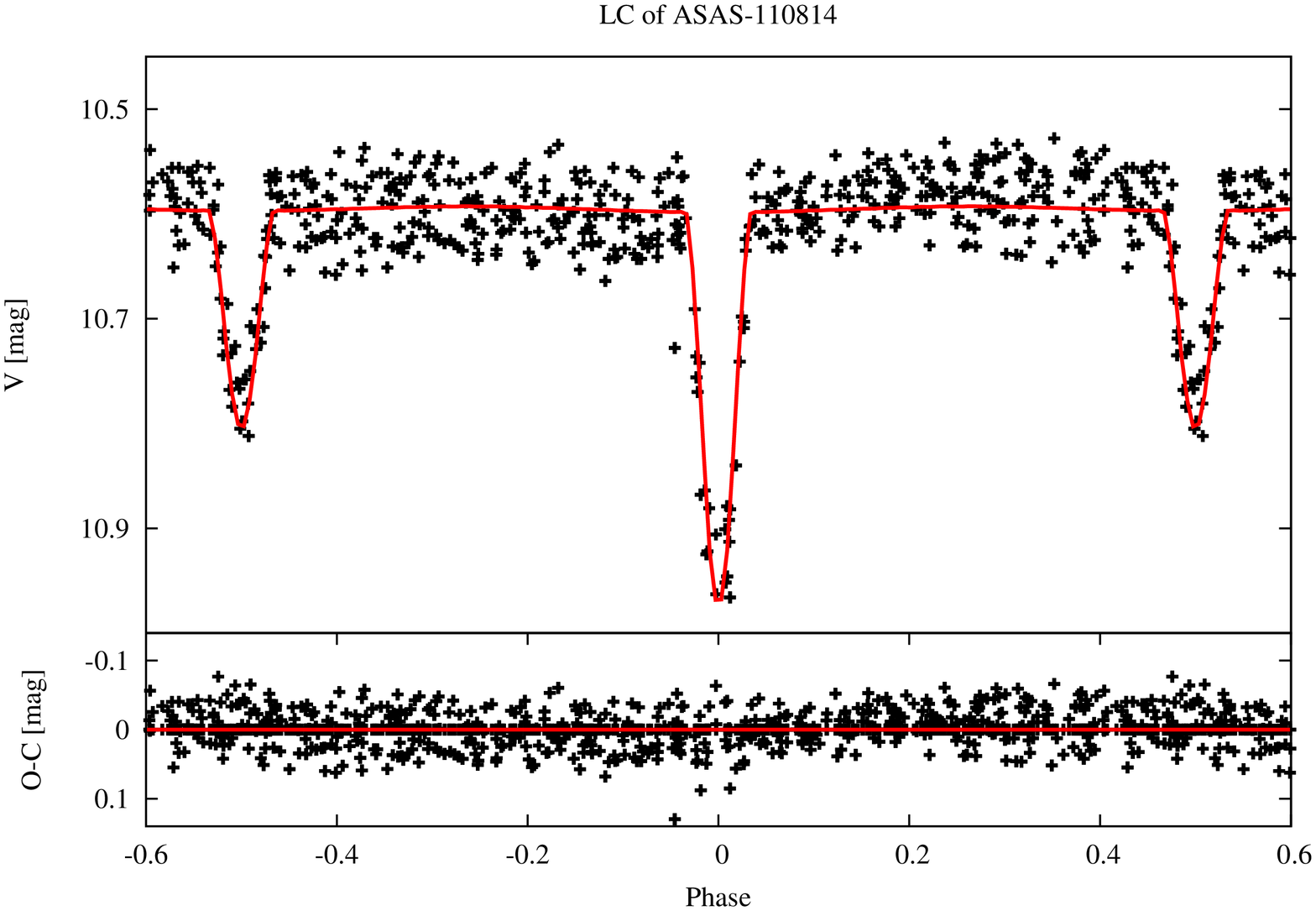}&
	\includegraphics[width=.27\textwidth, angle=-90]{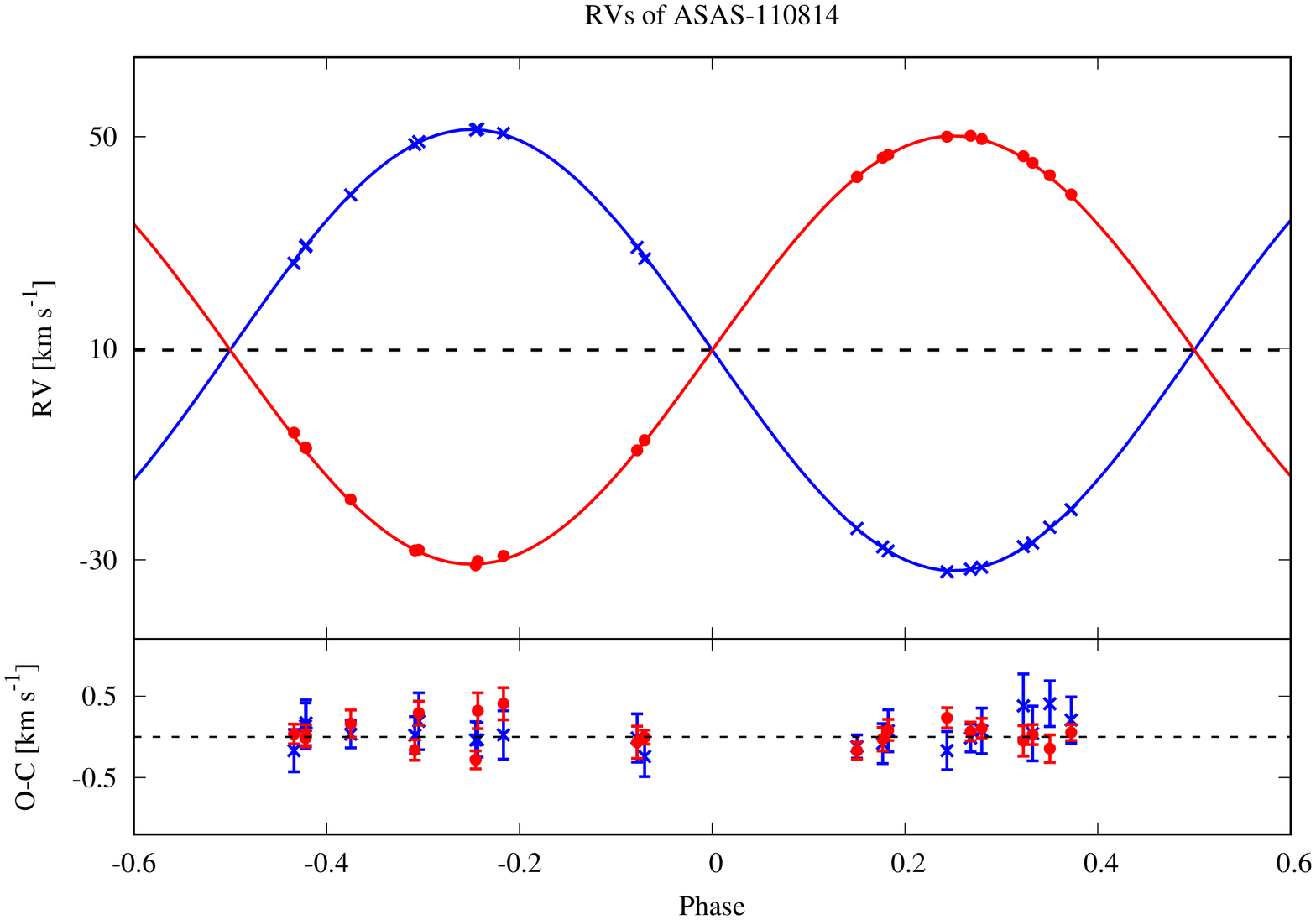}\\
	\includegraphics[width=.27\textwidth, angle=-90]{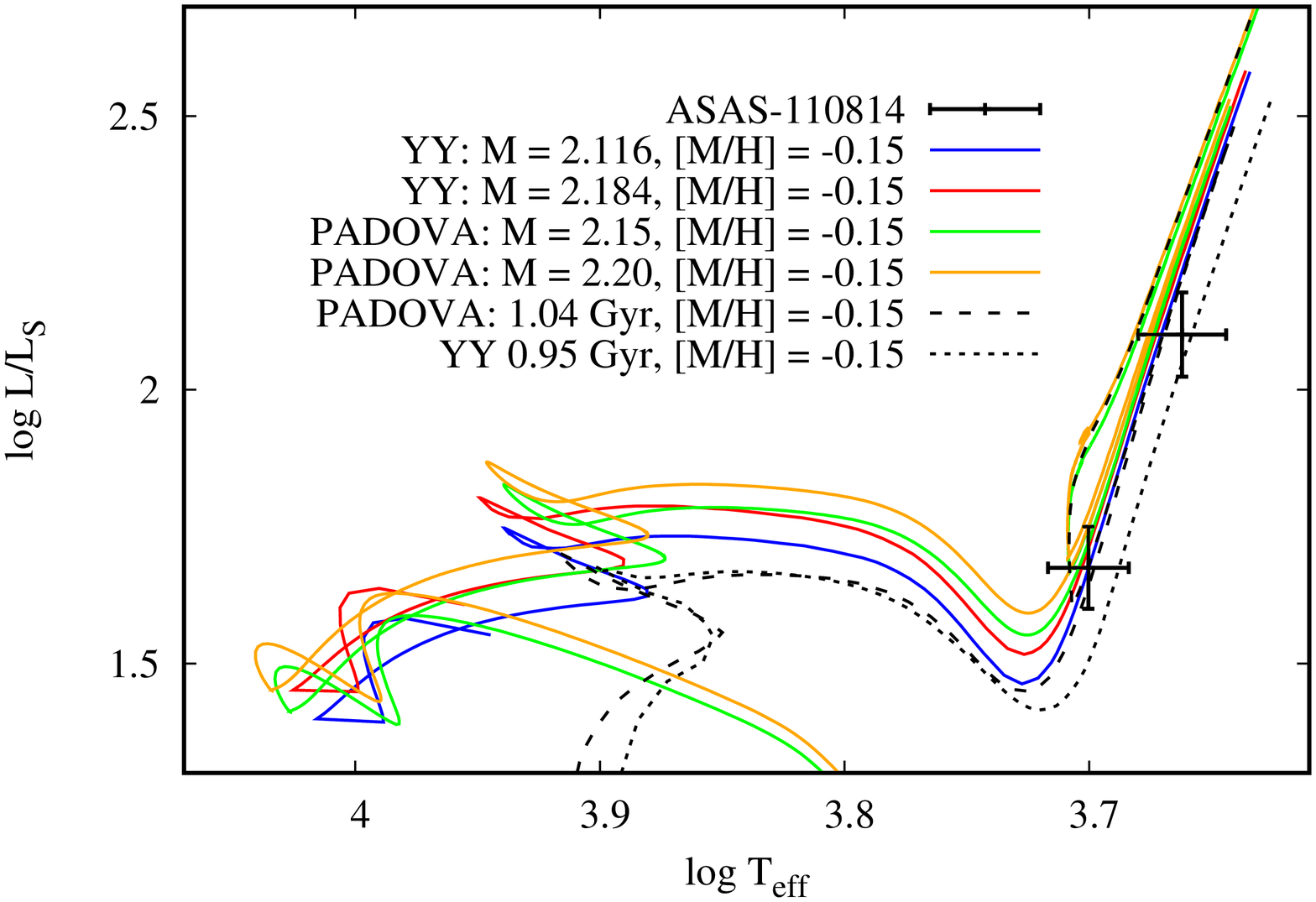}& 
	\includegraphics[width=.27\textwidth, angle=-90]{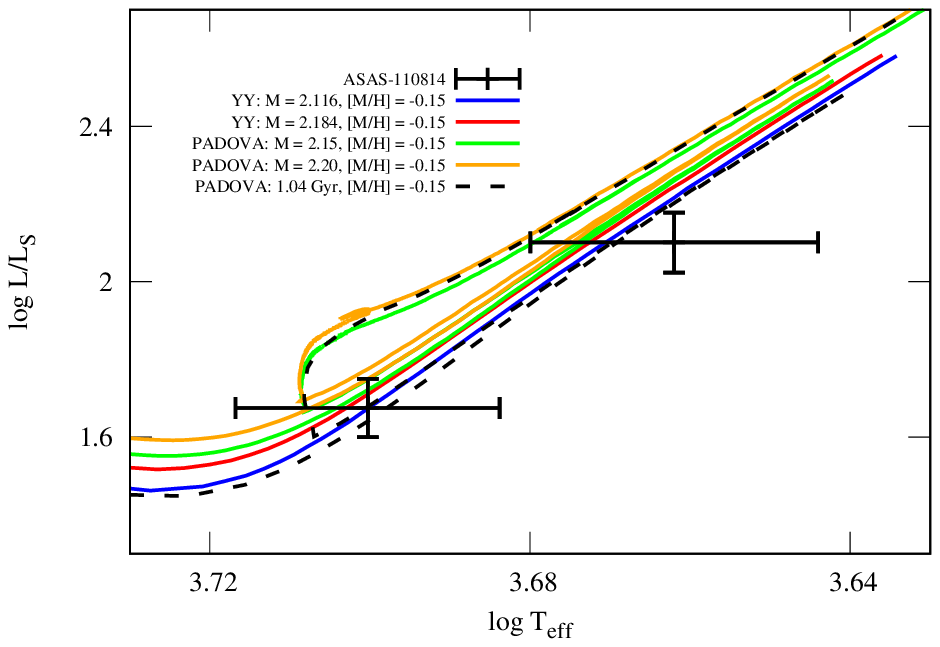}\\ 
	\end{tabular}
	\includegraphics[width=.3\textwidth, angle=-90]{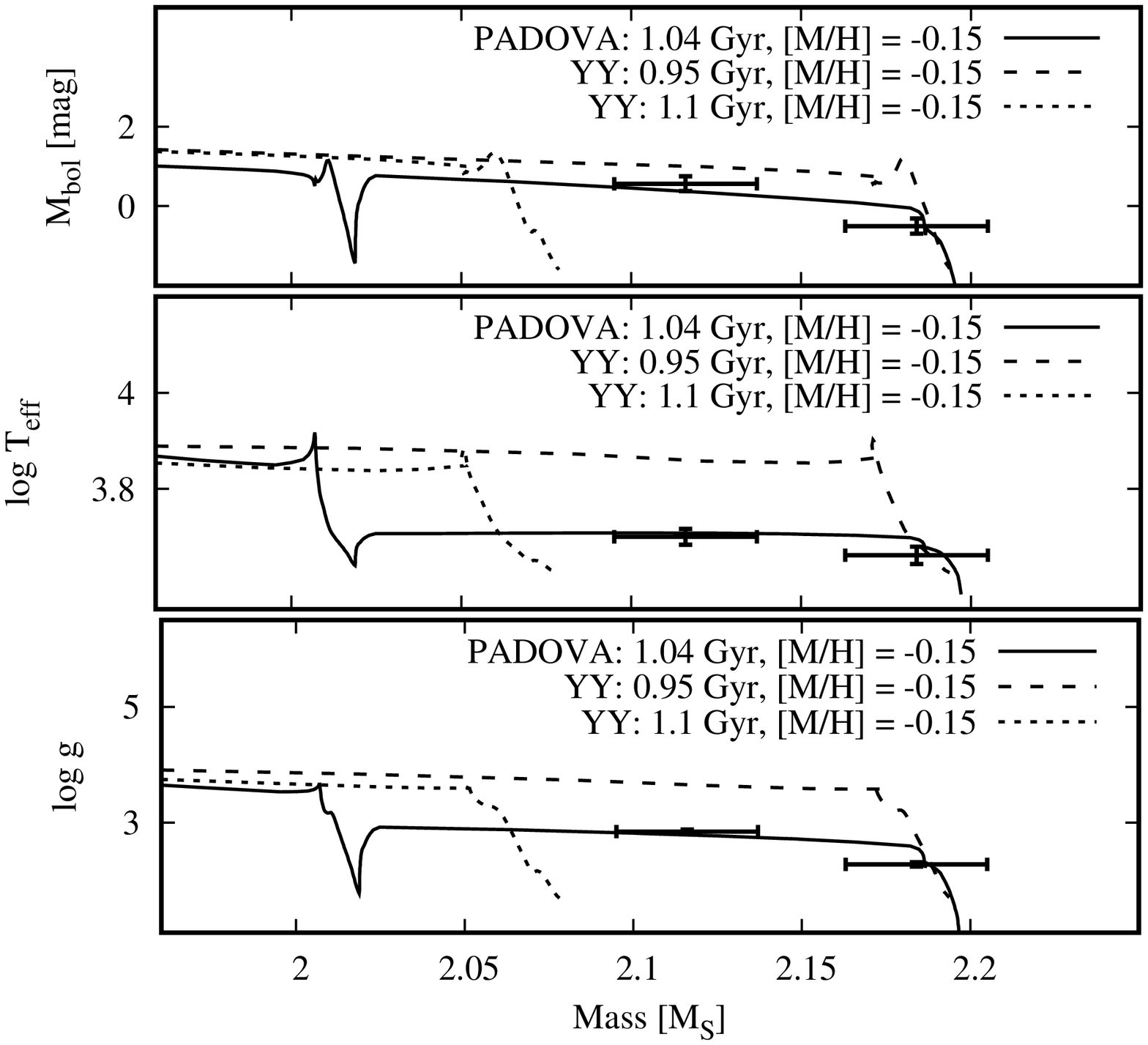}\\ 
	\caption{LC (ASAS-\textit{V}; left upper panel), RV curve (upper right), evolutionary tracks (middle panel; middle right is a zoom) and isochrones (lower panel) for A-110814. Blue colour represents primary component, red -- secondary. Green track represents  evolutionary tracks for $2.15~\Msun$ star, while yellow --  $2.20~\Msun$ and [M/H] = -0.15, calculated using PARSEC models.} 
	\label{110814_plots}
	\end{center}
\end{figure}

The analysis of A-110814 (see Fig.~\ref{110814_plots}) yielded masses and radii of $M_1 = 2.116 \pm 0.021~\Msun$, $M_2 = 2.184 \pm 0.021~\Msun$,  $R_1 = 9.105 \pm 0.367~\Rsun$, $R_2 = 17.759 \pm 0.550~\Rsun$.

Eleven CHIRON spectra were separated and analysed with \textsc{sme}. The inferred atmospheric parameters are: $T_\mathrm{eff1} = 5016 \pm 190$ K, $T_\mathrm{eff2} = 4950 \pm 195$ K, [M/H]$_1 = -0.10 \pm 0.11$, and [M/H]$_2 = -0.21 \pm 0.08$. The metallicity of the system was estimated by averaging those of the components: [M/H]$ = -0.15 \pm 0.11$.

Evolutionary tracks calculated for given masses and isochrones indicate that both components are in advanced stages of evolution. YY tracks imply that both stars are at the RG branch, but PARSEC isochrones indicate that both stars may be more evolved, reaching phases beyond YY models description. Approximate evolutionary tracks calculated with PARSEC models for $M = 2.15~\Msun$, $M = 2.20~\Msun$ and [M/H]$ = -0.15$ (no interpolator available to perform tracks calculation for precisely given masses and metallicity) do not exclude the scenario with components in more evolved stages (e.g. after the helium flash; see the zoom at the middle right panel of Fig.~\ref{110814_plots}). However we cannot prove this by studying  the tracks alone, as the errors of $T_\mathrm{eff}$ and $L$ are too large. As we failed in finding one isochrone from YY models fitting the data (see the lower panel of Fig.~\ref{110814_plots}, we conclude that the stars could indeed be more evolved. PARSEC isochrones (reaching the phases beyond YY models) entail that the components inhabit the red clump region, with the secondary already leaving it. The age of the system is estimated as $\sim 1.02 \substack{+0.1 \\ -0.1}$ Gyr (PARSEC).  

ASAS \textit{V}-band LC, RV curve, evolutionary tracks and isochrones for the systems are displayed in Fig.~\ref{110814_plots}.

\subsection{Main sequence pairs}

\subsubsection{A-062926}

\begin{figure}
	\begin{center}
	\begin{tabular}{rc}
	\includegraphics[width=.3\textwidth, angle=-90, scale =0.9]{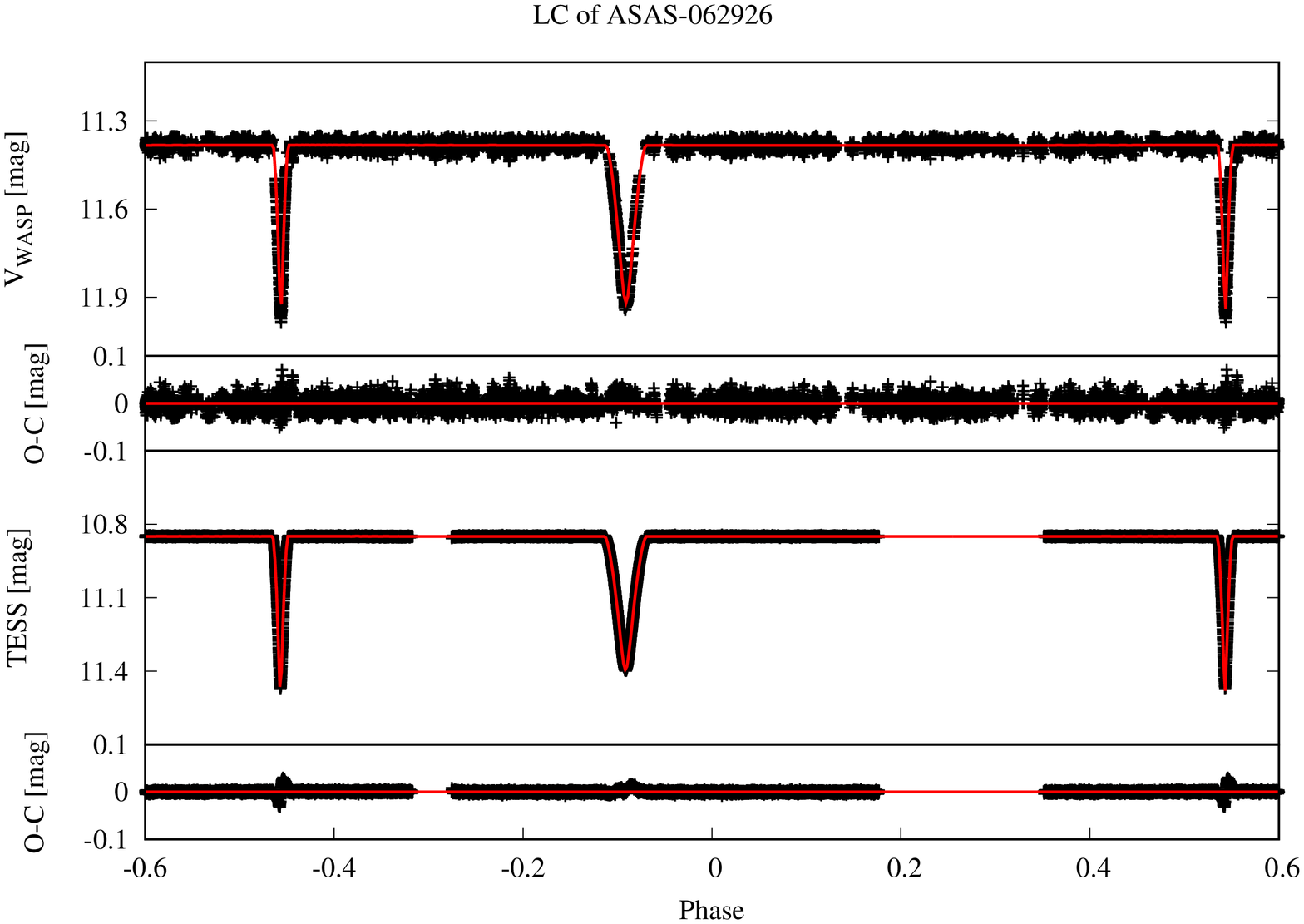}&
	\includegraphics[width=.27\textwidth, angle=-90]{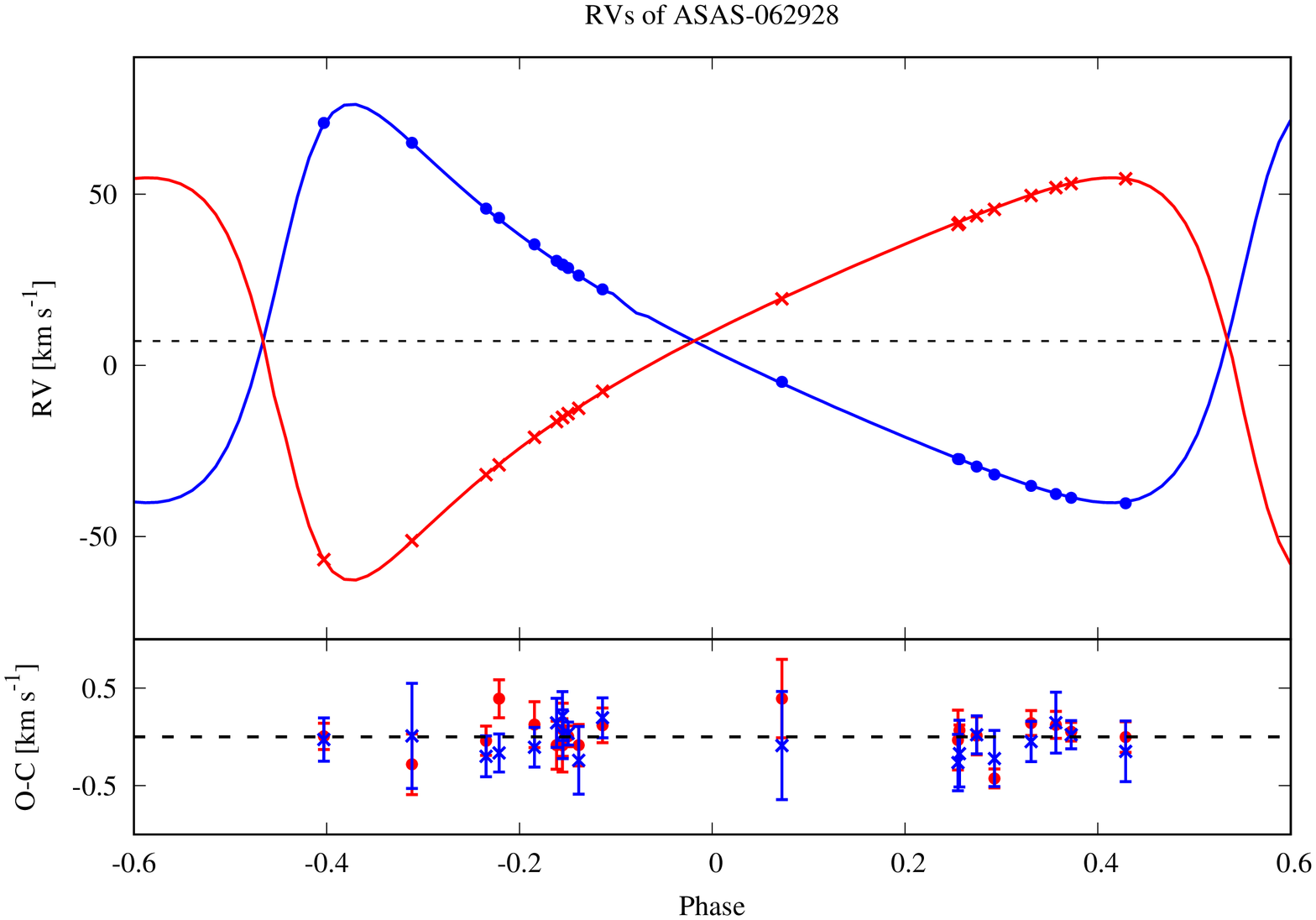}\\
	\includegraphics[width=.27\textwidth, angle=-90]{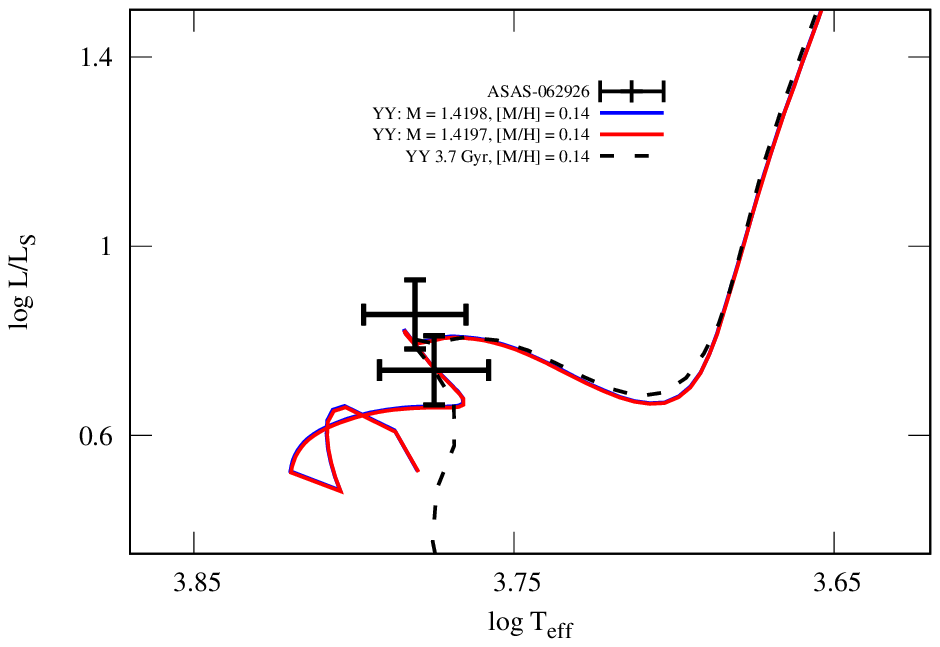}&
	\includegraphics[width=.27\textwidth, angle=-90]{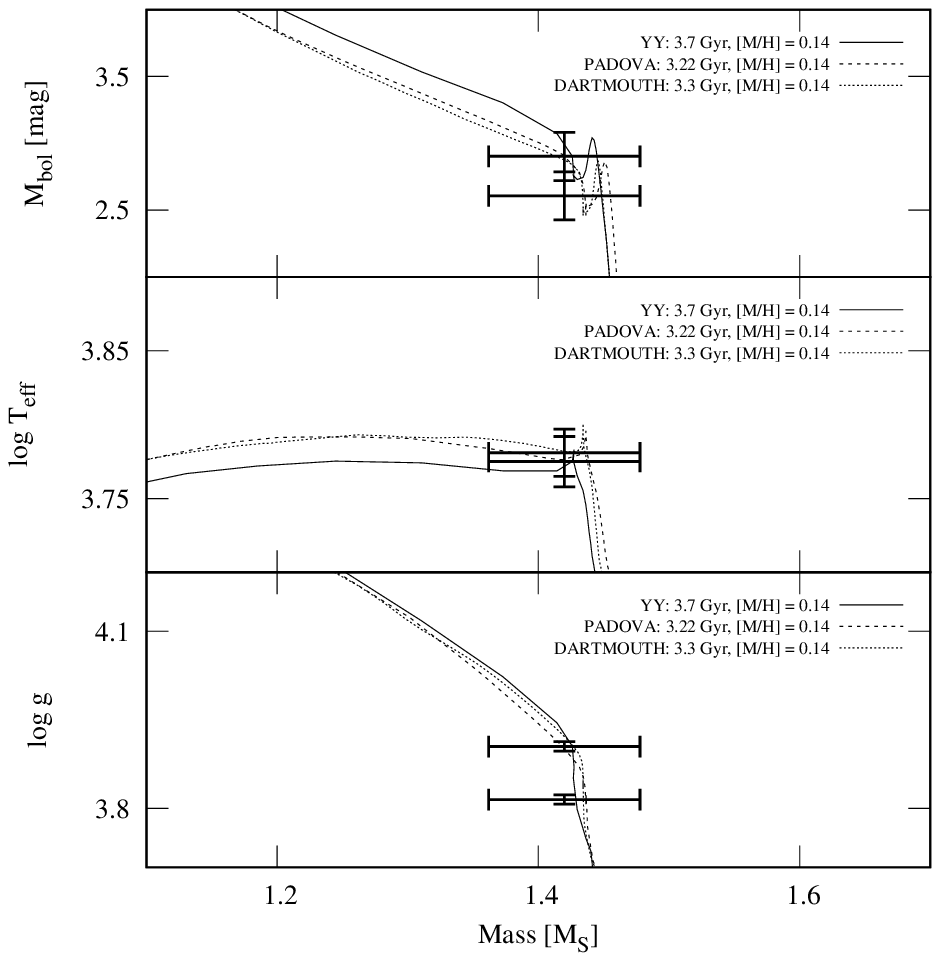}\\
	\end{tabular}
	\caption{LC (WASP, TESS; left upper panel), RV curve (upper right), evolutionary tracks (left lower) and isochrones (lower right) for A-062926. Blue colour represents primary component, red -- secondary.} 
	\label{062926_plots}
	\end{center}
\end{figure}

Our estimate for the masses and radii of the stars of A-062926 (see Fig.~\ref{062926_plots}) are $M_1 = 1.4198 \pm 0.0580~\Msun$, $M_2 = 1.4197 \pm 0.0580~\Msun$, $R_1 = 2.439 \pm 0.035~\Rsun$, and $R_2 = 2.201 \pm 0.034~\Rsun$. It is an equal-mass eccentric system with $e = 0.50 \pm 0.02$. See Tables~\ref{tab_orb} and~\ref{tab_par} for further details.

Notice, only the WASP data was analysed, as the dataset is richer than the relatively smaller ACVS.

Eleven of CHIRON spectra were separated to perform spectral analysis, yielding the following values of atmospheric parameters: $T_\mathrm{eff1} = 6050 \pm 230$ K, $T_\mathrm{eff2} = 5950 \pm 231$ K, [M/H]$_1 = 0.13 \pm 0.16$, and [M/H]$_2 = 0.15 \pm 0.18$. We adopted the mean of metallicities as the final value: [M/H]$ = 0.14 \pm 0.17$.  

Evolutionary tracks imply that the components are evolved main sequence (MS) stars close to TAMS. With a slightly larger primary, the isochrones do not exclude that the stars are already leaving the MS. The age of the system was estimated in the range $3.22-3.7 \substack{+0.31 \\ -0.32}$ Gyr (PARSEC and YY, respectively.)

TESS, WASP LCs, RV curve, evolutionary tracks and isochrones are presented in Fig \ref{062926_plots}.

\subsubsection{A-090232}

\begin{figure}
	\begin{center}
	\begin{tabular}{rc}
\includegraphics[width=.27\textwidth, angle=-90]{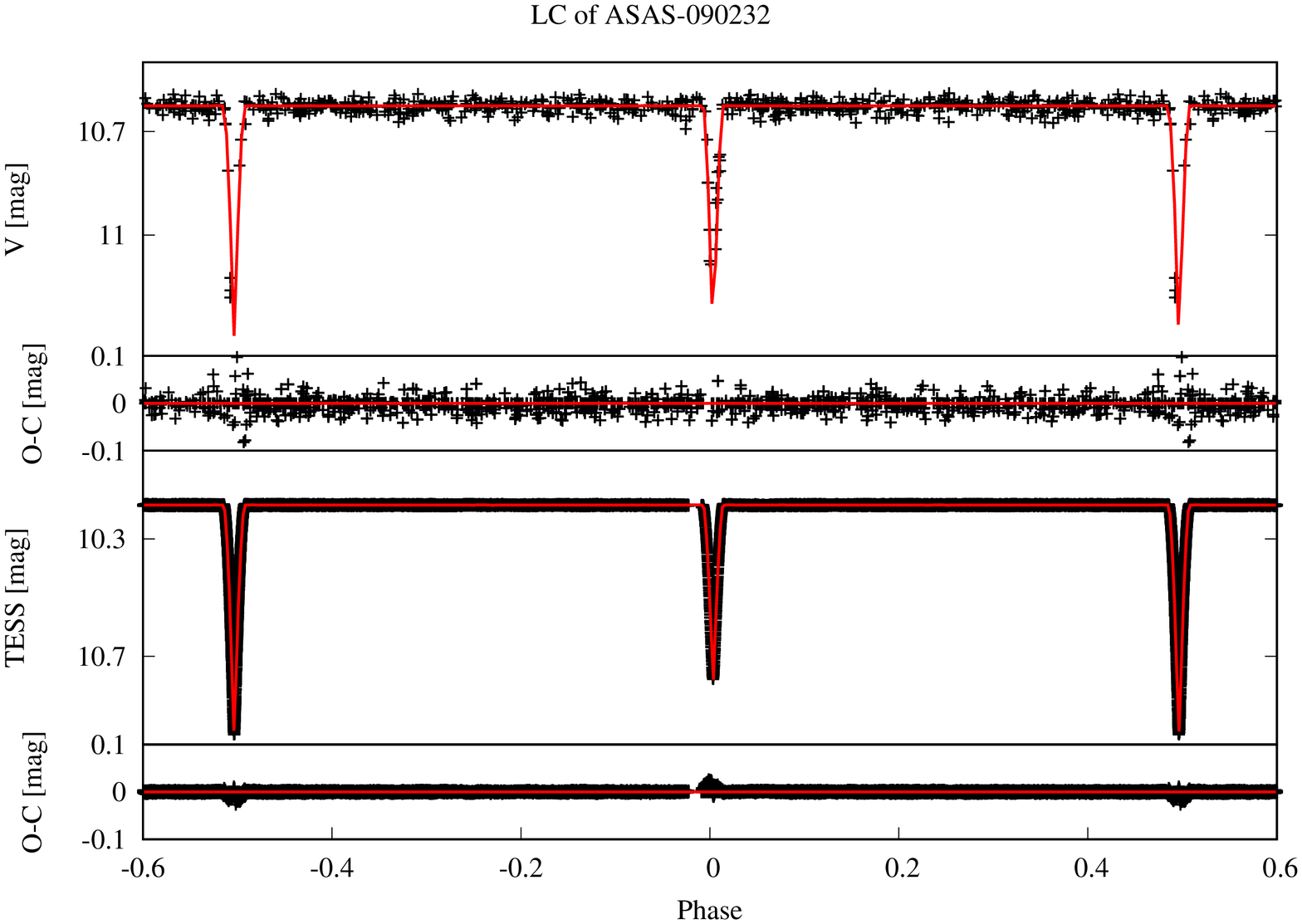}&
\includegraphics[width=.27\textwidth, angle=-90]{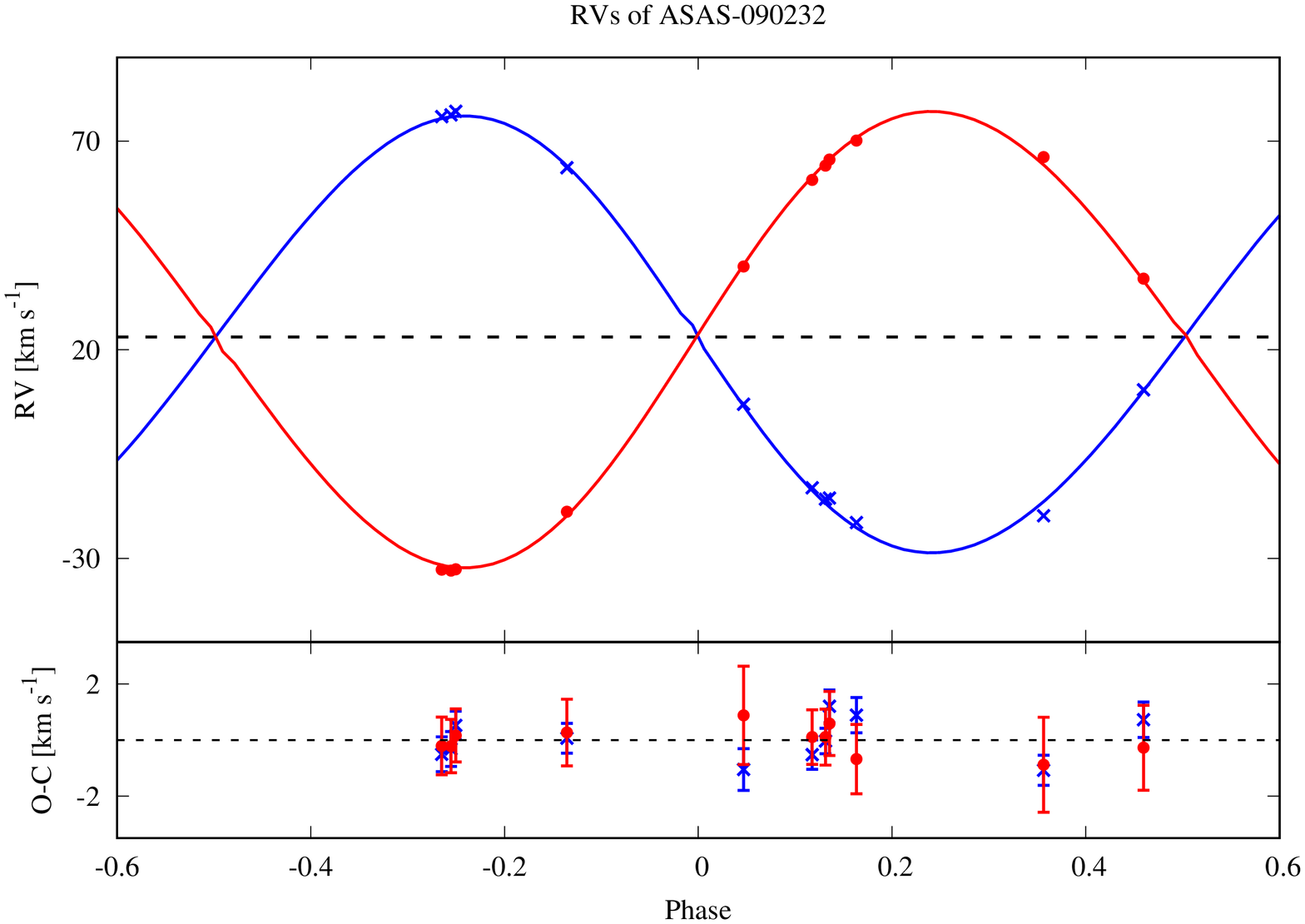}\\
\includegraphics[width=.27\textwidth, angle=-90]{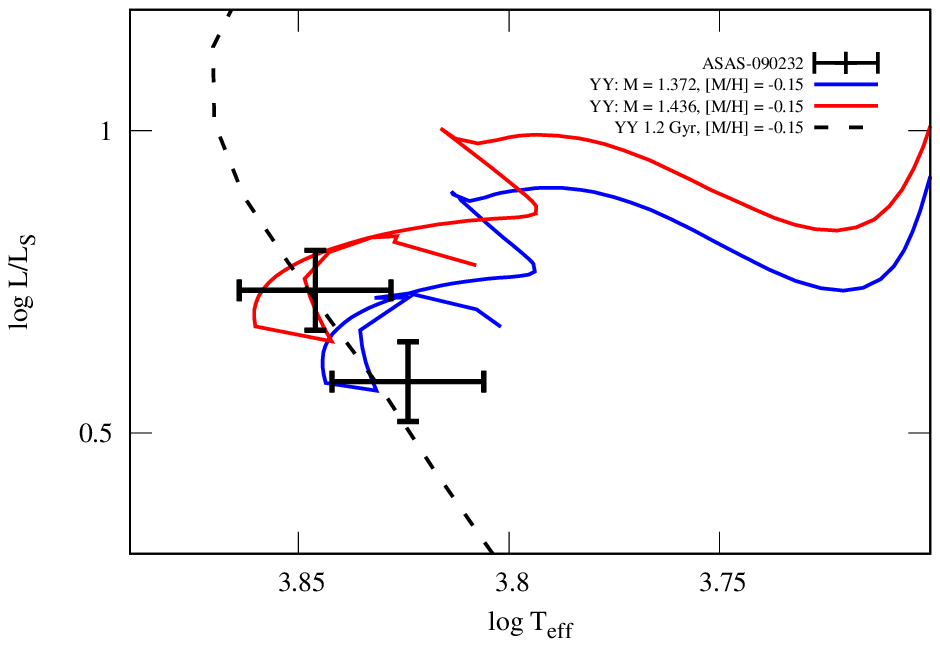}&
\includegraphics[width=.27\textwidth, angle=-90]{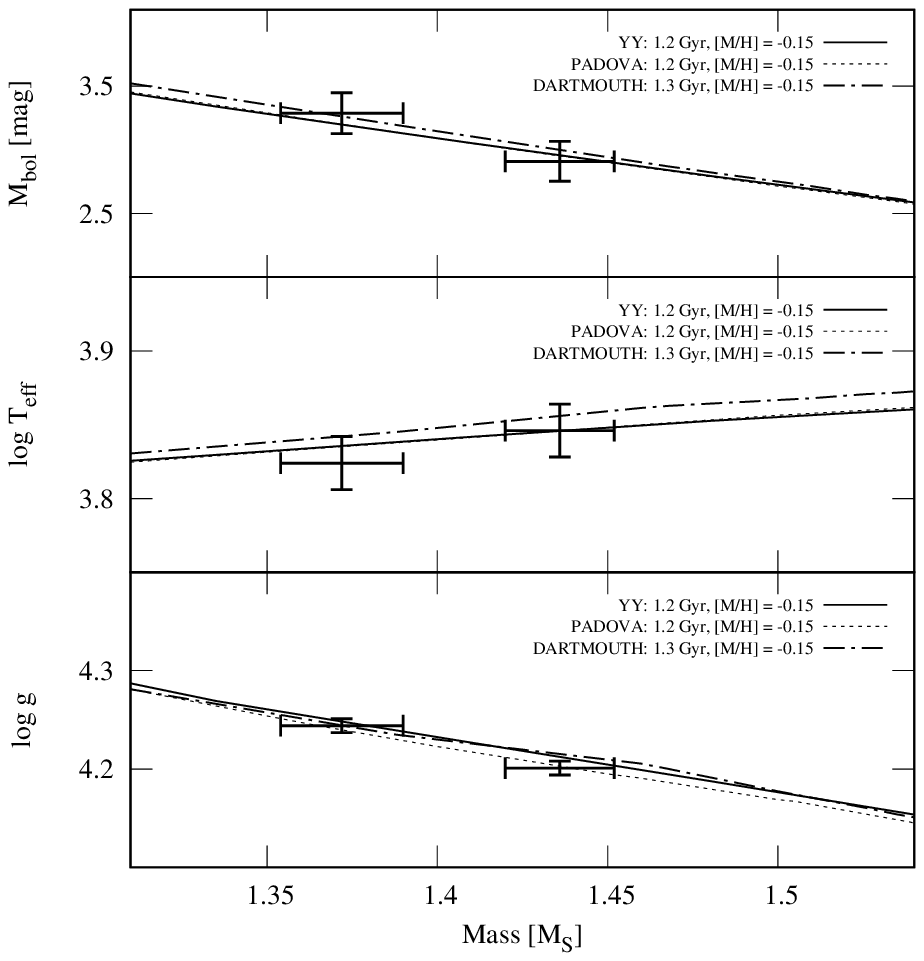}\\
	\end{tabular}
	\caption{LC (ASAS-\textit{V}, TESS; left upper panel), RV curve (upper right), evolutionary tracks (left lower) and isochrones (lower right) for A-090232. Blue colour represents primary component, red -- secondary.} 
	\label{090232_plots}
	\end{center}
\end{figure}

Our derived masses and radii for the stars in A-090232 (see Fig.~\ref{090232_plots}) are $M_1 = 1.436 \pm 0.016~\Msun$, $M_2 = 1.372 \pm 0.018~\Msun$, $R_1 = 1.392 \pm 0.007~\Rsun$, $R_2 = 1.527 \pm 0.007~\Rsun$. More orbital and physical parameters can be read in Tables~\ref{tab_orb} and~\ref{tab_par}.

Eleven CHIRON spectra were reconstructed and fed to spectral analysis, to obtain the following values for the atmospheric parameters: $T_\mathrm{eff1} = 7025 \pm 230$ K, $T_\mathrm{eff2} = 6680 \pm 230$ K, [M/H]$_1 = -0.17 \pm 0.12$, and [M/H]$_2 = -0.13 \pm 0.11$. Systemic metallicity was taken as the mean of those of the stars: [M/H]$ = -0.15 \pm 0.12$.   

YY evolutionary tracks indicate that the system consists of similar-mass stars still in the MS. Isochrones put the age of the system in the range $1.2-1.3  \substack{+0.15 \\ -0.1}$ Gyr (YY and Dartmouth, respectively).

TESS, ASAS \textit{V}-band LCs, RV curve, evolutionary tracks and isochrones for the systems are presented in Fig.~\ref{090232_plots}.

\subsection{System in different phases of evolution}

\subsubsection{A-051753}

\begin{figure}
	\begin{center}
	\begin{tabular}{rc}
	\includegraphics[width=.27\textwidth, angle=-90]{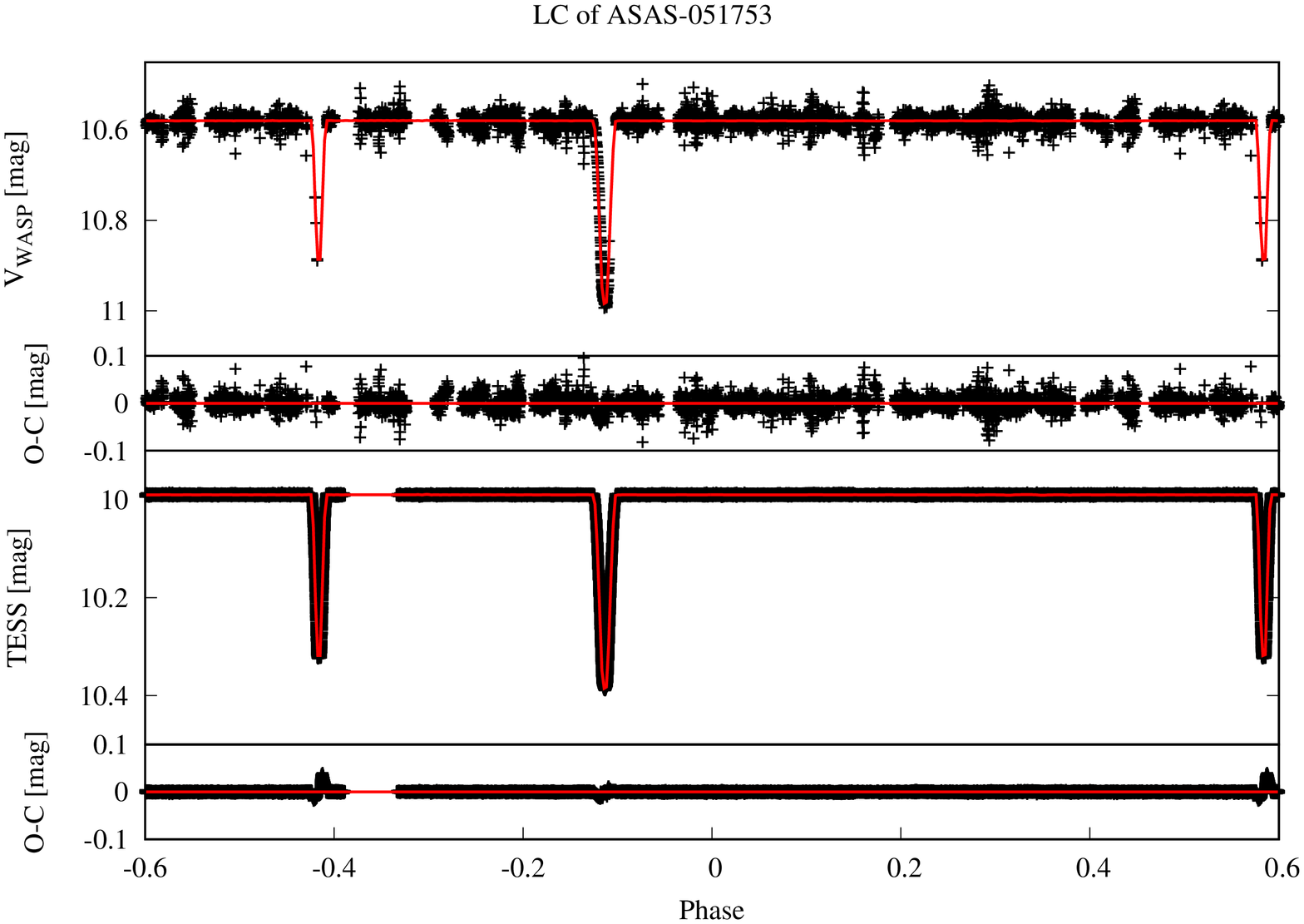}& 
	\includegraphics[width=.27\textwidth, angle=-90]{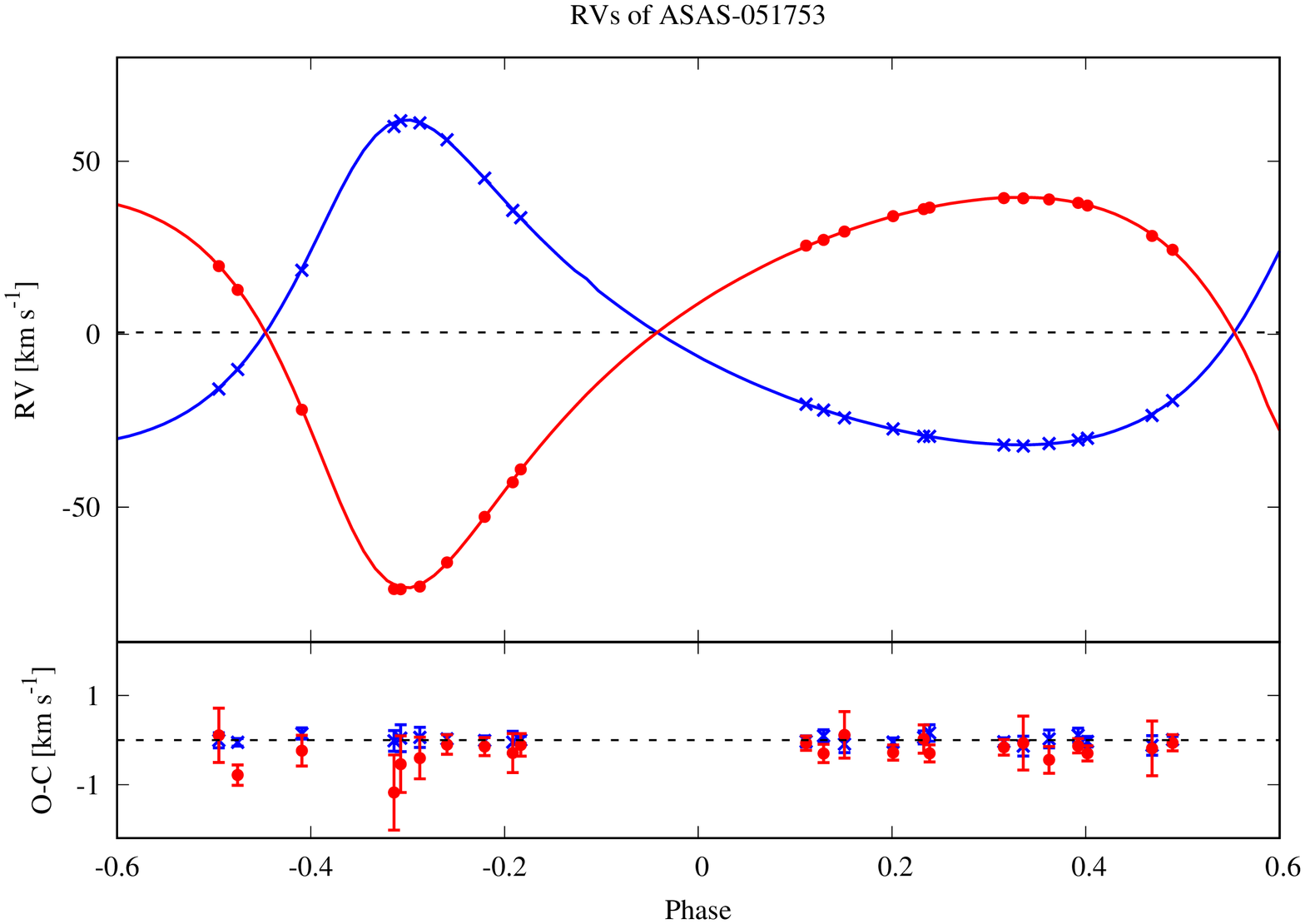}\\ 
	\includegraphics[width=.27\textwidth, angle=-90]{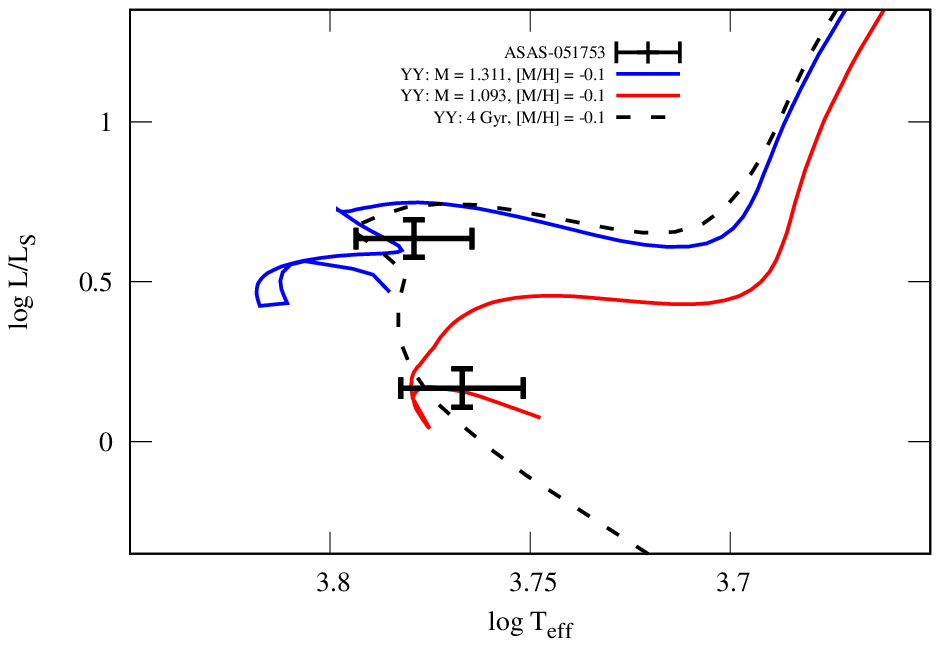}& 
 	\includegraphics[width=.27\textwidth, angle=-90]{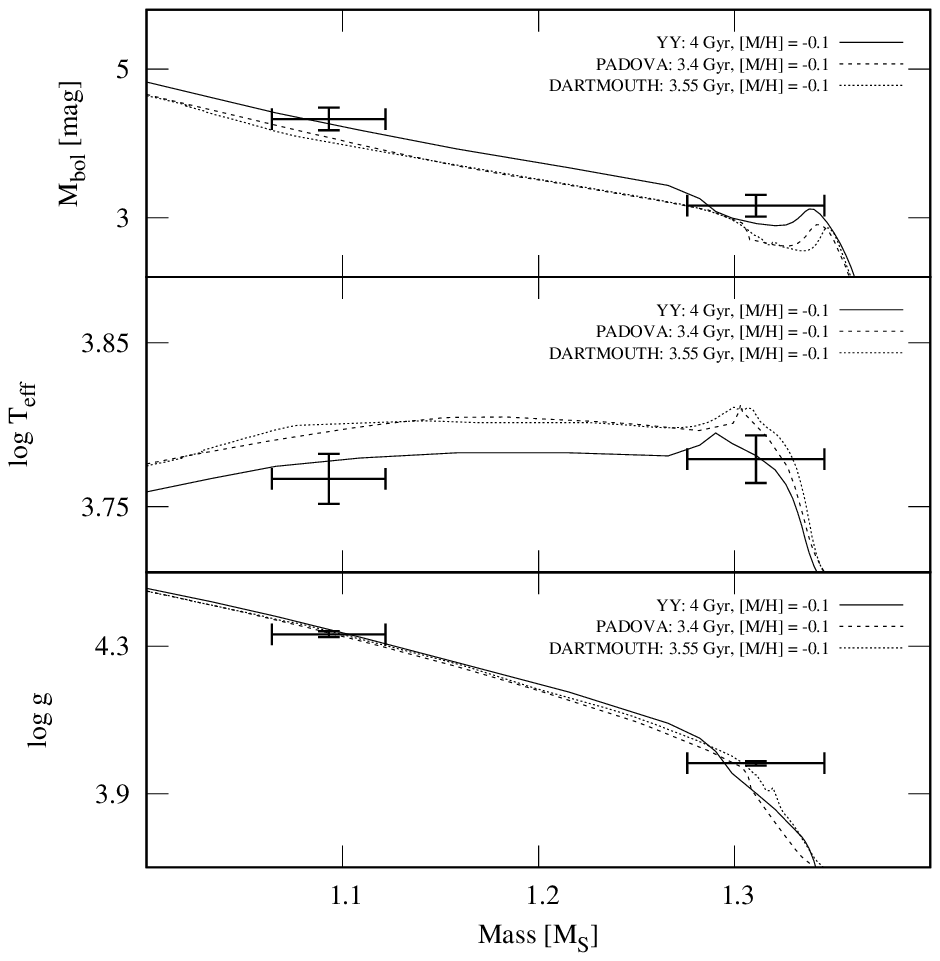}\\ 
	\end{tabular}
	\caption{LC (WASP, TESS; left upper panel), RV curve (right upper), evolutionary tracks (left lower) and isochrones (right lower) for A-051753. Blue colour represents primary component, red -- secondary.} 
	\label{051753_plots}
	\end{center}
\end{figure}

We estimated the individual masses and radii for A-051753 (see Fig.~\ref{051753_plots}) as $M_1 = 1.311 \pm 0.035~\Msun$, $M_2 = 1.093 \pm 0.029~\Msun$, $R_1 = 1.935 \pm 0.029~\Rsun$, and $R_2 = 1.179 \pm 0.014~\Rsun$. It is an eccentric system with $e = 0.369 \pm 0.02$. More orbital and physical parameters are presented in Tables~\ref{tab_orb} and~\ref{tab_par}. 

Due to the relatively low number of ACVS data with respect to WASP and TESS, we decided to analyse only the richer datasets. 

Ten spectra from CHIRON (fibre mode) were used for spectral separation, and the result was scaled and analysed with \textsc{sme}. The spectral analysis yielded $T_\mathrm{eff1} = 5980 \pm 205$ K, $T_\mathrm{eff2} = 5850 \pm 190$ K, [M/H]$_1 = -0.07 \pm 0.18$, [M/H]$_2 = -0.13 \pm 0.11$. The system metallicity (mean value of metallicities) yielded [M/H] $= -0.1 \pm 0.13$.
 
LC, RV curve, evolutionary tracks, and isochrones for A-051753 are shown in Fig.~\ref{051753_plots}. The stars are not in advanced stages of evolution: the secondary is a main sequence star, but the more massive primary could be an evolved MS star close to the TAMS, or even a sub-giant (the isochrones presented at the lower right panel of Fig.~\ref{051753_plots} favour such scenario). A-051753 could thus be a rare example of a system with components in slightly different evolutionary phases. Its evolutionary status resembles the slightly more massive TZ~For repeatedly used as a stringent test for stellar evolution models.

The age of the system was estimated as $3.4-4.0 \substack{+0.15 \\ -0.25}$ Gyr (PARSEC and YY models, respectively).

\subsection{V643 Ori}

\begin{figure}
	\begin{center}
	\begin{tabular}{cc}
	\includegraphics[width=.27\textwidth, angle=-90]{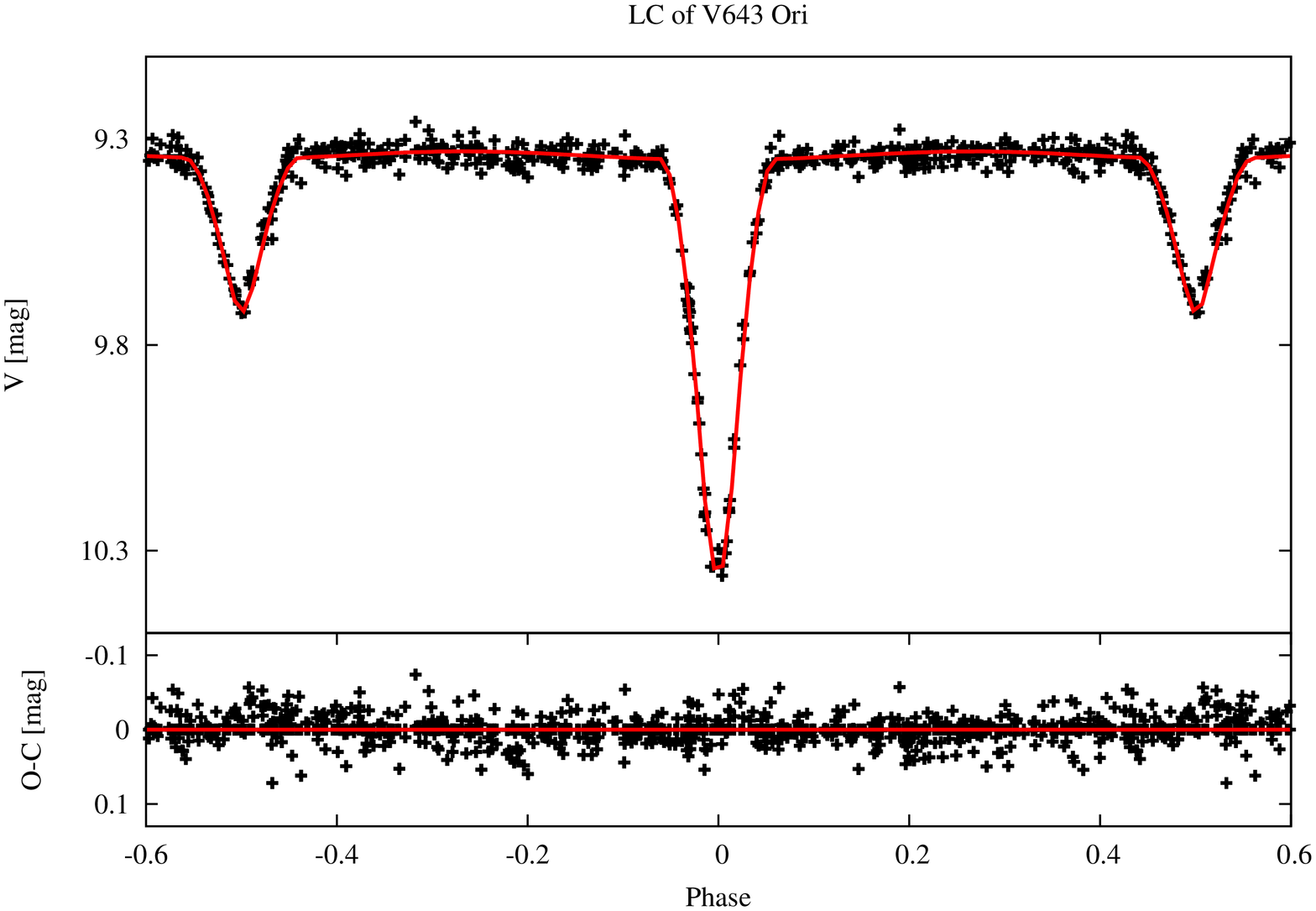}&
	\includegraphics[width=.27\textwidth, angle=-90]{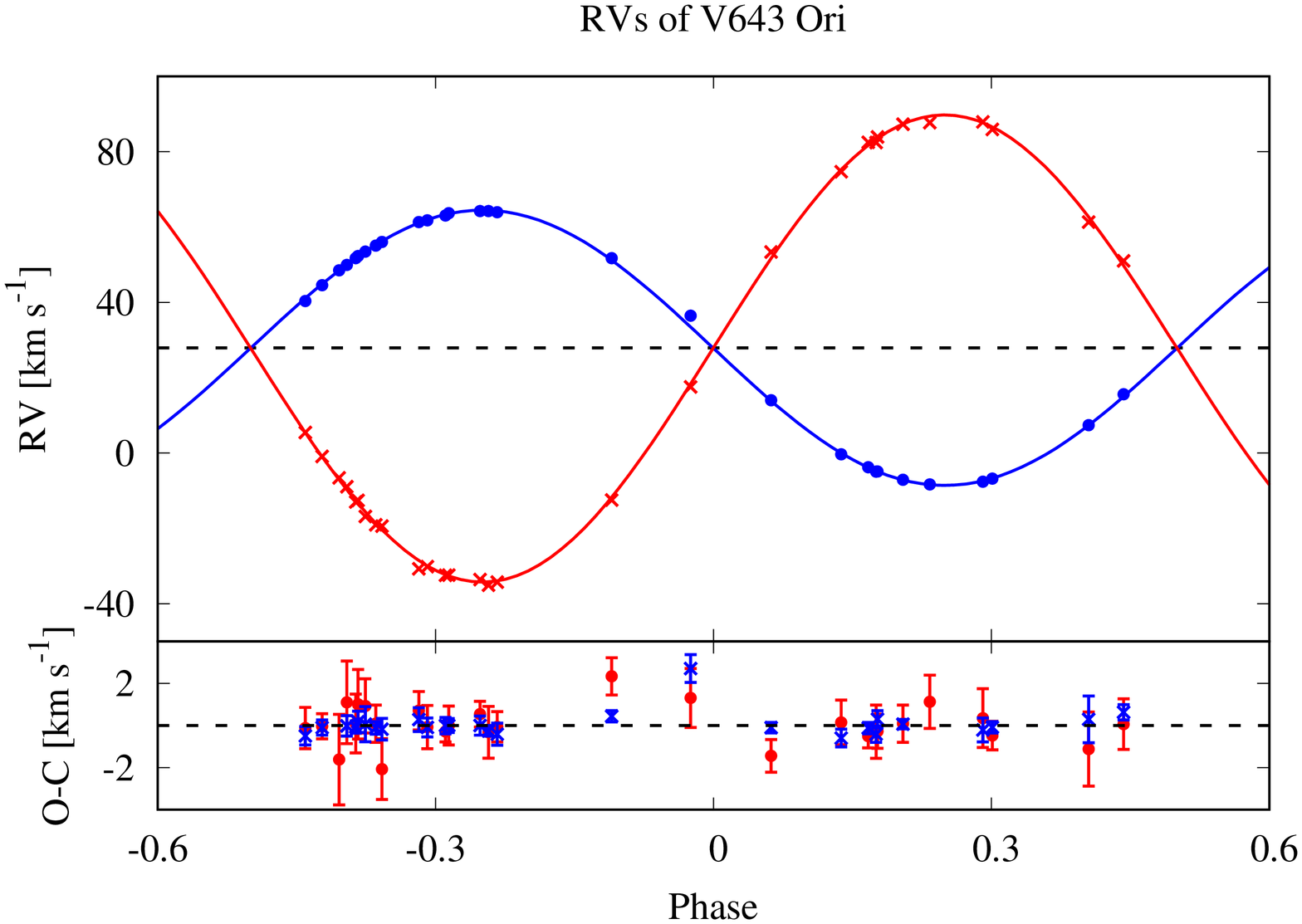}\\
	\includegraphics[width=.27\textwidth, angle=-90]{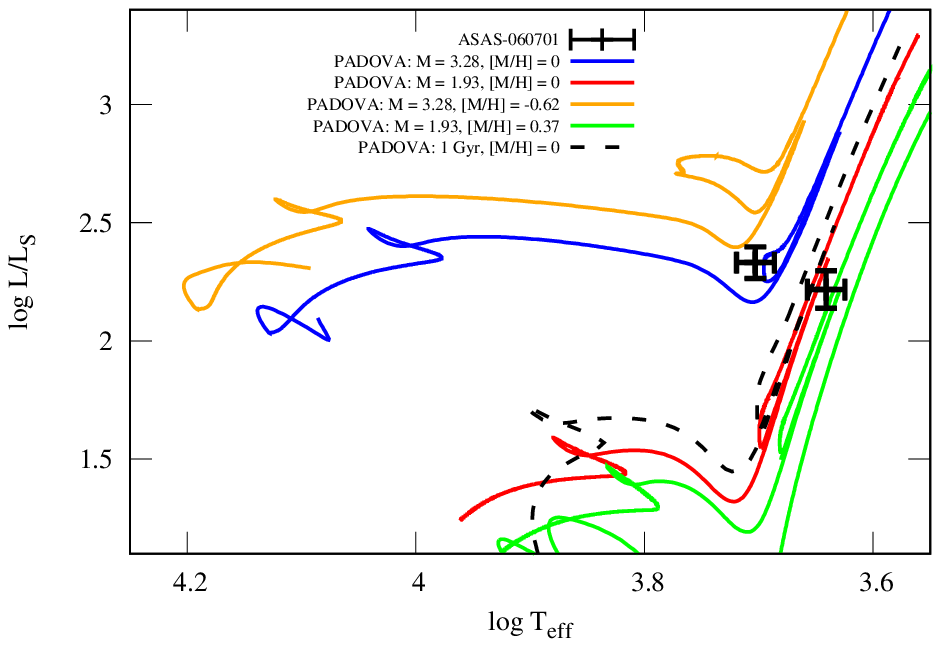}&
	\includegraphics[width=.27\textwidth, angle=-90]{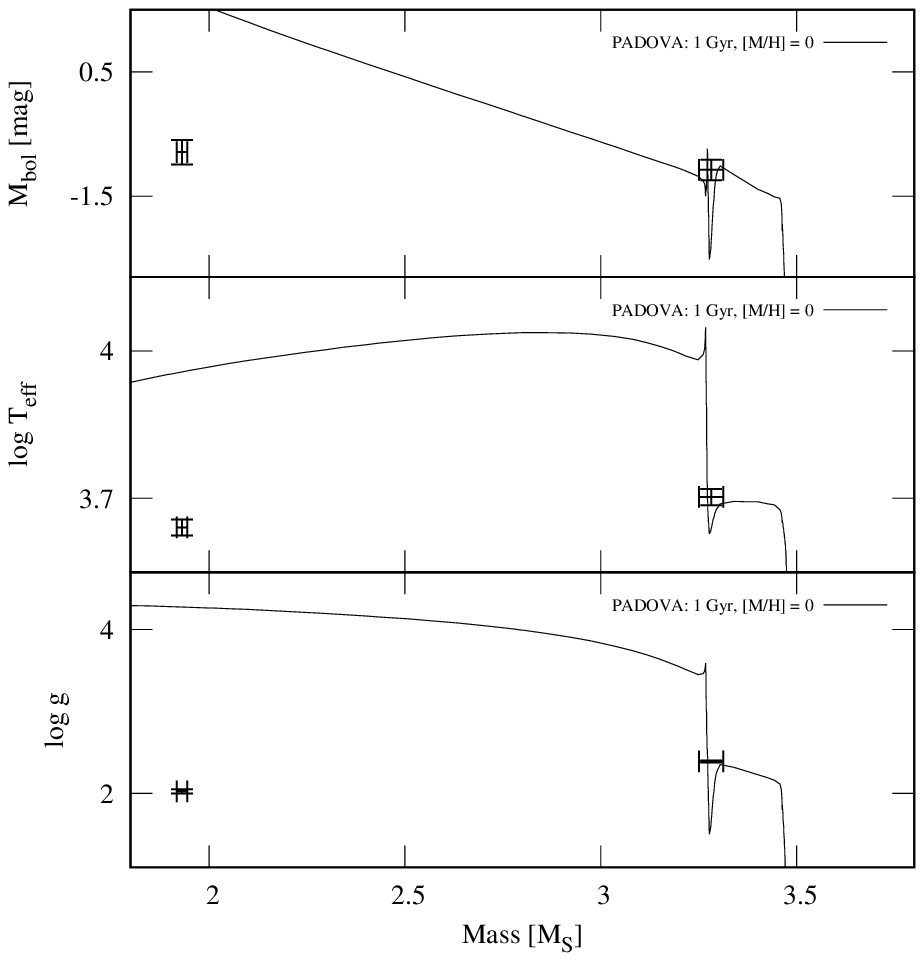}\\
	\end{tabular}
	\caption{LC (ASAS-\textit{V}; left upper panel), RV curve (upper right), evolutionary tracks (left lower) and isochrones (lower right) for V643~Ori. Blue colour represents primary component, red -- secondary. Green track represents an evolutionary track for  $1.95~\Msun$ star of ~[M/H] = 0.37, yellow -- star of $3.2~\Msun$ and ~[M/H] = -0.62, calculated with PARSEC models. For comparison we show also evolutionary tracks for solar metallicity [M/H] = 0 for the given masses.} 
	\label{060701_plots}
	\end{center}
\end{figure}

We derived masses and radii for the components of V643~Ori (see Fig.~\ref{060701_plots}) of $M_1 = 3.282 \pm 0.031~\Msun$, $M_2 = 1.931 \pm 0.014~\Msun$, $R_1 = 19.105 \pm 0.31~\Rsun$, and $R_2 = 22.272 \pm 0.62~\Rsun$. More derived data are available in Tables~\ref{tab_orb} and~\ref{tab_par}.

The number of spectra from the same spectrograph (fourteen) were enough to apply spectral separation and perform a spectral analysis. This yielded $T_\mathrm{eff1} = 5050 \pm 190$ K, $T_\mathrm{eff2} = 4380 \pm 190$ K, [M/H]$_1 = +0.37 \pm 0.11$, [M/H]$_2 = -0.62 \pm 0.11$.

The difference in metallicities for the stars made it problematic to estimate the final [M/H] of the system. Evolutionary tracks show that there is disagreement between the values of the parameters we obtained and theoretical models. The significant difference in the masses of the single stars (mass ratio $\sim 1.7$) prevents fitting a single isochrone to both, irrespective of the assumed [M/H] value ([M/H]$_1$, [M/H]$_2$, mean of values, or even solar [M/H] = 0). This fact upholds the peculiarity of the system and concurs with the description of V643~Ori given by \citet{egg06} and \citet{egg16}, who suggested that it might have a more sophisticated history -- possibly affected by stellar winds or where the primary might in turn be the result of a binary merger. Sec.~\ref{met} describes our narrative of the system evolution, supported by our findings. We believe that the problem of V643 Ori could be tackled using dedicated data of better quality and a more sophisticated close binary evolution analysis.

ASAS LC, RV curve, evolutionary tracks and isochrones are shown in Fig.~\ref{060701_plots}.

\subsection{Distance}

In order to calculate distances to the targets, we converted bolometric magnitudes into passband-dependent magnitudes using the corrections by \citet{bes98}. These yield the distance to the systems, when compared to apparent magnitudes (here in \textit{B}, \textit{V}, \textit{J}, \textit{H}, \textit{K} passbands from the Tycho-2 and 2MASS catalogues). To gauge the interstellar reddening, we took the value of the colour excess $E(B-V)$ from dust infrared emission maps of \citet{sch98} recalibrated by \citet{sch11} as an upper limit. $E(B-V)$ was further tuned to the value for which all five passband-dependent distances from \textsc{jktabsdim} agree. The adopted value of distance was taken as the average of these five values.

The inferred distances are compared with the GDR2 results in Table~\ref{tab_distance}. All our numbers are in relatively good agreement with those in GDR2, apart from the case of A-061016.
For a distant object as the latter, the error seems underestimated and if we take the most recurrent catalogue uncertainty instead, agreement is fully recovered.

\begin{table}
\begin{center}
\caption{Comparison of estimated targets distances with GDR2 data results.}
\label{tab_distance}
\begin{tabular}{ccc}
\hline
\hline
Object ID   & $d$ [pc] & $d_\mathrm{GDR2}$ [pc]\\
\hline
\rule{0pt}{3ex} A-051753 &    $365 \pm 31$  &    $383\substack{ +5 \\  -5}$ \\
\rule{0pt}{3ex} V643 Ori & $1\,163 \pm 119$ & $1\,229\substack{+78 \\ -69}$ \\
\rule{0pt}{3ex} A-061016 & $1\,003 \pm 115$ & $1\,185\substack{+40 \\ -37}$ \\
\rule{0pt}{3ex} A-062926 &    $708 \pm 48$  &    $768\substack{+17 \\ -16}$ \\
\rule{0pt}{3ex} A-065114 &    $785 \pm 115$ &    $860\substack{+30 \\ -28}$ \\
\rule{0pt}{3ex} A-090232 &    $480 \pm 68$  &    $440\substack{ +7 \\  -6}$ \\
\rule{0pt}{3ex} A-110814 & $1\,315 \pm 126$ & $1\,440\substack{+78 \\ -70}$ \\
\hline
\end{tabular}
\end{center}
\end{table}

\section{Discussion}

Several problems encountered during our investigation are worth mentioning. Below, we discuss the subjects of brightness ratio, metallicity, and activity.

\subsection{Brightness ratio}
\label{br}

We used the BR values in the wavelength range of our spectroscopic observations instead of the values determined from the LC analysis, as the TESS bandpass is wide and more sensitive to red wavelengths. We found it appropriate to rescale the reconstructed spectra from a given spectral range with the BR value determined from the same wavelength range. As there can be some degeneracy between the ratio of the radii and the BR determined from the LC for partially eclipsing binaries, we found it compelling to use the BR from spectroscopy.
We compared the BR values from the TODCOR and light curve analyses (choosing the dataset of better quality) and found them in agreement (see Tab.~\ref{tab_orb}, \ref{tab_par}).

\begin{figure}
\includegraphics[width=.35\textwidth, angle=-90, scale = 0.97]{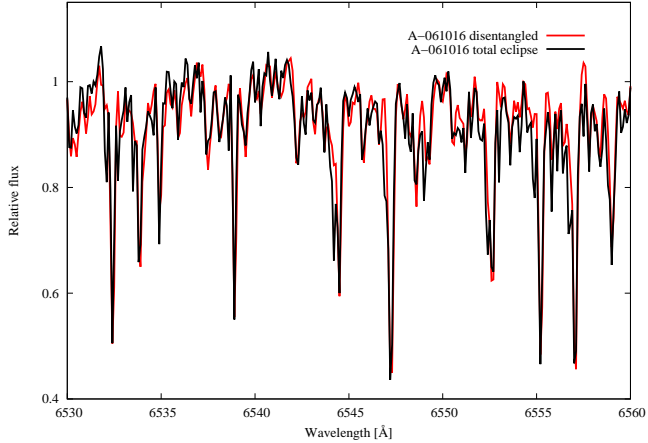}
	\caption{A-061016 secondary spectrum taken during the total eclipse (black) compared with reconstructed and rescaled CORALIE spectrum (red).} 
	\label{sme_comp_061016}
\end{figure}

 Figure \ref{sme_comp_061016} presents the spectrum of A-061016 secondary taken during the total eclipse compared with the separated and rescaled spectrum from the CORALIE set. The agreement confirms that the value of the BR from the TODCOR analysis used to rescale the reconstructed spectra was correct. This justifies our approach at calculating the final values of the BR by averaging the values from every order of all spectra. As an example, Fig.~\ref{sp_comp_sme} presents the separated and scaled spectra of the stars in A-051753 compared with synthetic models from the calculated atmospheric parameters.

\subsection{Metallicity} 
\label{met}
 
Another problematic issue is the difference of metallicities from the separated spectra of stars in the same system. In most cases, averaging yielded a trustworthy value in agreement with the models, but the case of V643~Ori shows how elusive the matter can be. No single isochrone could be fitted to the parameters we obtained, neither taking the system metallicity as that of one of the stars, nor taking the average. Cases are known where the two components have different metallicities, but these are mainly close semi-detached binary systems \citep{pav12}, where mass exchange occurs between closely-separated stars. The puzzle confirms that V643~Ori is an intriguing target, whose components could have not evolved as single detached stars. \citet{egg06} suggested that one of the mechanism at the root of mass transfer could be binary-enhanced stellar winds (BESW): the author claims that the primary experiences mass loss, reaching the phase of Roche lobe filling, just before the helium flash. Albeit fascinating, the process does not explain the high mass ratio of the components. An alternative process sees the primary as the result of the merger of two stars: V643~Ori could thus be a former triple with final components of masses and periods of $2.0 + 1.34~\Msun$ and $P = 1.5$ d + $1.93~\Msun$ and $P = 52$ d, respectively. A dedicated spectroscopic campaign and more sophisticated analysis, including the close binary evolution, is needed to confirm any of these hypotheses.

\subsection{Activity}

Two of our systems show out-of-eclipse brightness variations that could be caused by spots on the stellar surface -- A-065114 and A-110814. The LC phase coverage for these objects is too poor to perform spots modelling, but the out-of-eclipse brightness variation over time indicate seasonal changes in stellar temperature. In order to inspect chromospheric activity we have checked the emission in the H$\alpha$ and Ca \textsc{II} H and K spectral band (available only for part of the data), but none was found. A-065114 is listed in the ROSAT catalogue of X-ray sources and described in literature as a plausible example of a target showing coronal activity, making the absence of emissions and deficiency of chromospheric activity somewhat surprising. While it is not clear which components of these systems show spots, following the results of our previous studies \citep{rat16} we suspect that only the larger, cooler components of these systems are spotted. However, we have no confirmation of this fact from our data.

\section{Conclusions}

We have investigated seven double-lined DEBs from the ACVS using CHRION spectroscopic data (followed by more spectroscopic observations in some cases) and ASAS/WASP/TESS photometry. Tomographic separation was applied to perform spectral analysis and determine atmospheric parameters of the stars. Even if five out of the seven targets consist of nearly equal-mass systems, the components are in slightly different phases of evolution, as the majority of them has already left the MS. At this stage of evolution, stars of similar mass offer a wide range of radii and effective temperatures \citep[see e.g.,][]{hel15}. Therefore, a precise determination of masses and metallicities is required to constrain their age and exact evolutionary stage. The survey resulted in the studies of two MS-MS systems, one possible MS-subgiant pair, three RG-RG, and one possible merger. For two RG-RG systems (A-065114 and A-110814) we have discussed more advanced phases of evolution (red clump stars or stars on the AGB). Systems with possible MS-subgiant stars (A-051753) could be very informative cases of stellar pair with components of different masses and thus dissimilar evolutionary stages. Such objects are extremely important to improve codes modelling stellar structure and evolution.      

High quality photometric (TESS) and spectroscopic data let us obtain the components masses and radii with precision better than $3\%$ for four systems (A-051753, V643~Ori, A-061016, A-090232), making them reliable test-beds for evolutionary models. With a dedicated spectroscopic campaign and advanced close binary studies V643~Ori is a good specimen to investigate the evolution of two stars in advanced phase and to test merging scenario with binary evolution codes. In order to better outline the phases of evolution of the stars in A-061016, more advanced evolution codes like \textsc{mesa} \citep{pax11} should be used. A-051753 in turn enrich the sparse sample of well defined binaries with components in slightly different evolutionary stages. All systems constitute a valuable contribution to DEBCat, filling the gap of stars leaving TAMS and adding up to the sample of RGs, while A-090232 is an example of well-characterised MS stellar pair.

\section*{Acknowledgements}

We would like to thank A. Mazierska for her help in calculating radial velocities from the HARPS spectra and M. Armano for the valuable comments. In this research, we have used the Aladin service, operated at CDS, Strasbourg, France, the SAO/NASA Astrophysics Data System Abstract Service and the Mikulski Archive for Space Telescopes operated by the Space Telescope Science Institute. This work includes data collected by the TESS mission. Funding for the TESS mission is provided by the NASA Explorer Program. MR, KGH, MK, PM acknowledge the support provided by the Polish National Science Center (NCN) through grants 2015/16/S/ST9/00461, 2016/21/B/ST9/01613, 2017/27/B/ST9/02727, and 2016/21/B/ST9/01126, respectively.

\section*{Data availability}
The data underlying this article will be shared on reasonable request to the corresponding author.

\bsp	




\begin{landscape}

\begin{table}
\begin{center}
\caption{RMS of fitting of photometric ($\sigma_\mathrm{LC}$) and spectroscopic ($\sigma_\mathrm{RV}$) data, and mean formal error of photometric (A-\textit{V} -- ASAS-\textit{V}, A-\textit{I} -- ASAS-\textit{I}, W -- WASP, T -- TESS) measurements ($\overline{\sigma}_\mathrm{LC}$) for the studied systems.}
\label{tab_rms}
\begin{tabular}{ccccc}
\hline
\hline
Object ID & $\sigma_\mathrm{LC}$ [mag] & $\overline{\sigma}_\mathrm{LC}$ [mag] & $\sigma_\mathrm{RV1}$ [km~s$^{-1}$]& $\sigma_\mathrm{RV2}$ [km~s$^{-1}$]\\
\hline
A-051753 & 0.013(W), 0.001(T) & 0.013(W), 0.001(T) & 0.19 & 0.29 \\
V643 Ori & 0.024 & 0.024 & 0.79 & 0.93 \\
A-061016 & 0.022(A-\textit{V}), 0.028(A-\textit{I}), 0.021(W) & 0.022(A-\textit{V}), 0.028(A-\textit{I}), 0.021(W) & 0.31 & 0.20\\ 
A-062926 & 0.012(W), 0.002(T) & 0.012(W), 0.002(T) & 0.15 & 0.17\\ 
A-065114 & 0.031 & 0.032 & 0.33 & 1.00\\
A-090232 & 0.037(A-\textit{V}), 0.002(T)  & 0.037(A-\textit{V}), 0.002(T) & 0.77 & 0.52\\
A-110814 & 0.039 & 0.039 & 0.17 & 0.17\\
\hline
\end{tabular}
\end{center}
\end{table}

\begin{table}
\caption{Orbital elements and spectroscopic brightness ratio of A-051753, V643 Ori, A-061016, A-062926, A-065114, A-090232, and A-110814. $T_{0}$ is given as JD-2450000 d, $\gamma_{2} - \gamma_{1}$ indicates the difference between components centre-of-mass RVs.}
\label{tab_orb}
\resizebox{1.35 \textwidth}{!}{%
\begin{tabular}{lrlrlrlrlrlrlrl}
\hline
\hline
Parameter & \multicolumn{2}{c}{A-051753} & \multicolumn{2}{c}{V643 Ori} & \multicolumn{2}{c}{A-061016}& \multicolumn{2}{c}{A-062926} & \multicolumn{2}{c}{A-065114}& \multicolumn{2}{c}{A-090232} & \multicolumn{2}{c}{A-110814}\\
\hline
$P$ [d] &   $26.13021$ & $\pm 0.00041$ & $52.42130$ & $\pm 0.00072$ &$199.8569$ & $\pm 0.0039$ & $26.38163$ & $\pm 0.00031$ & $43.4869$ & $\pm 0.0021$ & $20.821914$ & $\pm 0.000042$ & $75.2173$ & $\pm 0.0027$ \\
$T_{0}$ &  $1914.096770$ & $\pm 0.000051$ & $1911.543$ & $\pm 0.020$ & $2128.61$ & $\pm 0.58$ &  $1916.33230$ & $\pm 0.00010$ & $2413.014$ & $\pm 0.083$ & $8524.778580$ & $\pm 0.000040$ &  $1886.631$ & $\pm 0.069$\\
$K_{1}$ [km~s$^{-1}$]  & $47.03$ & $\pm 0.18$ & $36.529$ & $\pm 0.091 $ & $24.263$ & $\pm 0.072$ & $58.58$ & $\pm 0.18$ & $46.901$ & $\pm 0.094$ & $53.33$ & $\pm 0.33 $ &  $41.723$ & $\pm 0.048$ \\
$K_{2}$ [km~s$^{-1}$] & $56.39$ & $\pm 0.17$ & $62.08$ & $\pm 0.25 $ & $24.176$ & $\pm 0.064$ & $58.59$ & $\pm 0.17$ & $47.86$ & $\pm 0.22$ & $55.85$ & $\pm 0.26 $ & $40.431$ & $\pm 0.056$ \\
$\gamma_{1}$ [km~s$^{-1}$]&  $0.16$ & $\pm 0.29$ & $27.98$ & $\pm 0.13$ & $37.61$ & $\pm 0.11$  & $7.048$ & $\pm 0.079$ & $22.09$ & $\pm 0.10$ & $23.12$ & $\pm 0.29$ & $9.818$ & $\pm 0.057$\\
$\gamma_{2} - \gamma_{1}$ [km~s$^{-1}$]&  $0.0$ & $(*)$ & $-0.55$ & $\pm 0.43$ & $0.08$ & $\pm 0.11$ & $0.0$ & $(*)$ & $0.0$ & $(*)$ & $0.16$ & $\pm 0.33$ & $-0.371$ & $\pm 0.055$\\
$e$ & $0.367$ & $\pm 0.020$ & $0.0$ & $(*)$ & $0.127$ & $\pm 0.021$ &$ 0.503 $ & $\pm 0.020$ & $0.0$ & $(*)$ & $0.0$ & $(*)$ & $ 0.058 $ & $\pm 0.026$\\
$i$ & $89.741$ & $\pm 0.010$ & $88.5$ & $\pm 1.0$ & $88.02$ & $\pm 0.12$ & $89.251$ & $\pm 0.021$ & $81.1$ & $\pm 1.7$ & $89.81$ & $\pm 0.03$ & $89.81$ & $\pm 1.16$\\
$a$ [R$_\odot$] & $49.70$ & $\pm 0.44$ & $102.17$ & $\pm 0.28$ & $189.84$ & $\pm 0.63$ & $52.70$ & $\pm 0.72$ & $82.42$ & $\pm 0.43$ & $44.75$ & $\pm 0.17 $ & $121.89$ & $\pm 0.37$\\
$\omega [^{\circ}]$ & $327.81$ & $\pm 0.82$ & -- & & $174.13$ & $\pm 0.63$ & $291.72$ & $\pm 0.22$ & -- & & -- & & $271.1$ & $\pm 1.8$\\
$BR_\mathrm{sp}$ & 0.39 & & 2.56 & & 2.53 & & 0.74 & & 2.17 & & 0.73 & & 3.85\\
\hline
\end{tabular}
}
\end{table}
\end{landscape}

\begin{landscape}
\begin{table}
\caption{Physical parameters and photometric brightness ratio of A-051753, V643 Ori, A-061016, A-062926, A-065114, A-090232, and A-110814. Values of effective temperature $T_\mathrm{eff}$, metallicity [M/H] and rotational velocity $v_\mathrm{rot}$ are taken from spectral analyses of reconstructed spectra.}
\label{tab_par}
\resizebox{1.35 \textwidth}{!}{%
\begin{tabular}{lrlrlrlrlrlrlrl}
\hline
\hline
Parameter & \multicolumn{2}{c}{A-051753} & \multicolumn{2}{c}{V643 Ori} & \multicolumn{2}{c}{A-061016}& \multicolumn{2}{c}{A-062926} & \multicolumn{2}{c}{A-065114}& \multicolumn{2}{c}{A-090232} & \multicolumn{2}{c}{A-110814}\\
\hline
$M_{1}$ [$\Msun$] &  $1.311$ & $\pm 0.035$ & $3.282$ & $\pm 0.031$ & $1.148$ & $\pm 0.031$ & $1.4198$ & $\pm 0.058$ & $2.009$ & $\pm 0.033$ & $1.436$ & $\pm 0.016$ &  $2.116$ & $\pm 0.019$\\
$M_{2}$ [$\Msun$] & $1.093$ & $\pm 0.029$ & $1.931$ & $\pm 0.014$ & $1.152$ & $\pm 0.012$ &  $1.4197$ & $\pm 0.058$ & $1.968$ & $\pm 0.030$ & $1.372$ & $\pm 0.018$ &  $2.183$ & $\pm 0.020$\\
$R_{1}$ [$\Rsun$]  & $1.935$ & $\pm 0.020$ & $19.11$ & $\pm 0.31$ & $12.34$ & $\pm0.19$ & $2.441$ & $\pm 0.092$ & $9.6$ & $\pm 2.8$ & $1.527$ & $\pm 0.007$ & $9.11$ & $\pm 0.37$ \\
$R_{2}$ [$\Rsun$]  & $1.181$ & $\pm 0.014$ & $22.27$ & $\pm 0.62$  &  $33.32$ & $\pm 0.39$ &  $2.201$ & $\pm 0.034$ & $17.2$ & $\pm 1.5$ & $1.392$ & $\pm 0.007$ & $17.76$ & $\pm 0.55$\\
$\log g_{1}$ [cm~s$^{-2}$]  & $3.982$ & $\pm 0.006$ & $2.391$ & $\pm 0.014$  & $2.315$ & $\pm 0.014$ & $3.815$ & $\pm 0.007$ &$2.78$ & $\pm 0.26$ & $4.201$ & $\pm 0.006$ & $2.844$ & $\pm 0.035$\\
$\log g_{2}$ [cm~s$^{-2}$]  & $4.332$ & $\pm 0.008$ & $2.028$ & $\pm 0.024$ & $1.454$ & $\pm 0.010$ &  $3.905$ & $\pm 0.008$ & $2.259$ & $\pm 0.075$ & $4.244$ & $\pm 0.007$ & $2.278$ & $\pm 0.027$\\
$v_\mathrm{rot1}$ [km~s$^{-1}$] &  $3.6$ & $\pm 3.0$ & $19.27$ & $\pm 0.82$ & $6.08$ & $\pm 0.98$  & $14.7$ & $\pm 1.2$ & $14.22$ & $\pm 0.82$ & $4.1$ & $\pm 2.5$ & $11.32$ & $\pm 0.92$\\
$v_\mathrm{rot2}$ [km~s$^{-1}$] &  $2.1$ & $\pm 1.0$ & $25.3$ & $\pm 1.1$ &$7.1$ & $\pm 1.0$  & $15.3$ & $\pm 1.5$ & $25.13$ & $\pm 0.91$ & $3.9$ & $\pm 2.0$ & $16.60$ & $\pm 0.71$\\
$T_\mathrm{eff1}$ [K] &  $5980$ & $\pm 205$ & $5050$ & $\pm 190$ & $4290$ & $\pm 190$ &  $6050$ & $\pm 230$ & $4920$ & $\pm 203$ & $7025$ & $\pm 230$ & $5016$ & $\pm 190$\\
$T_\mathrm{eff2}$ [K] &  $5850$ & $\pm 190$ & $4380$ & $\pm 195$ & $3830$ & $\pm 186$  & $5950$ & $\pm 231$ & $4550$ & $\pm 190$ & $6680$ & $\pm 230$ & $4590$ & $\pm 195$\\
$\mathrm{[M/H]_{1}}$ & $-0.07/-0.1$&$\pm 0.18/0.13$&$0.37$&$\pm 0.13$&$0.09/0.08$&$\pm 0.10$&$0.13/0.14$&$\pm 0.17$&$-0.16/-0.13$&$\pm 0.16$&$-0.15/-0.17$&$\pm 0.12$&$-0.10/-0.15$&$\pm 0.11$\\
$\mathrm{[M/H]_{2}}$ & $-0.13/-0.1$&$\pm 0.13$&$-0.62$&$\pm 0.11$&$0.07/0.08$&$\pm 0.10$&$0.15/0.14$&$\pm 0.18$&$-0.10/-0.13$&$\pm 0.13$&$-0.15/-0.13$&$\pm 0.13$&$-0.20/-0.15$&$\pm 0.08$\\
$\log L_{1}$ [$\Lsun$] & $0.635$ & $\pm 0.059$ & $2.331$ & $\pm 0.067$ & $1.667$ & $\pm 0.074$& $0.856$ & $\pm 0.073$ & $1.68$ & $\pm 0.27$ & $0.736$ & $\pm 0.062$ & $1.675$ & $\pm 0.075$\\
$\log L_{2}$ [$\Lsun$] &  $0.168$ & $\pm 0.060$ & $2.217$ & $\pm 0.079$ & $2.319$ & $\pm 0.092$ & $0.738$ & $\pm 0.074$ & $2.060$ & $\pm 0.11$ & $0.585$ & $\pm 0.065$ & $2.278$ & $\pm 0.079$\\
$M_\mathrm{bol1}$ [mag] & $3.16$ & $\pm 0.15$ & $-1.08$ & $\pm 0.17 $ & $0.58$ & $\pm 0.19$ & $2.61$ & $\pm 0.17$ & $0.54$ & $\pm 0.68$ & $2.91$ & $\pm 0.16$ & $0.56$ & $\pm 0.19$\\
$M_\mathrm{bol2}$ [mag] &$4.33$ & $\pm 0.15$ & $-0.79$ & $\pm 0.20$ & $-1.05$ & $\pm 0.23$ & $2.90$ & $\pm 0.17$ & $-0.39$ & $\pm 0.26$ & $3.29$ & $\pm 0.16$ & $-0.50$ & $\pm 0.19$ \\
$BR_\mathrm{ph}$ & 0.35 & & 2.98 & & 2.72 & & 0.78 & & 2.63 & & 0.72 & & 4.37 & \\
\hline
\end{tabular}
}
\end{table}

\end{landscape}

\onecolumn

\twocolumn



\appendix

\section{RV measurements for A-051753, V643 Ori, A-061016, A-062926, A-065114, A-090232, and A-110814 systems}

The section includes Tables \ref{RV_table_051753}--\ref{RV_table_110814} with RV measurements, final RV errors, O-Cs, exposure times for each spectrum, SNR and instrument specifications for both components of the selected systems. The used telescopes/spectrographs are as follows: CTIO/CH = CTIO 1.5-m/CHIRON ( f -- fibre mode, s -- slicer mode) EUL/C = Euler/CORALIE, ESO/F = MPG/ESO 2.2-m/FEROS, ESO/H = ESO 3.6m/HARPS, AAT/U = AAT 3.9m/UCLES. SNR stands for a signal-to-noise ratio per collapsed spectral pixel at $\lambda$=5\,500~\AA.

\onecolumn
\begin{table}
\caption{RV measurements for A-051753.}
\centering
\begin{tabular}{c c c c c c c c c c}
\hline 
BJD-2450000 & $RV_{1}$ & $\sigma_\mathrm{RV_{1}}$ & $O-C_{1}$ & $RV_{2}$ & $\sigma_\mathrm{RV_{2}}$ & $O-C_{2}$ & $T_\mathrm{exp}$ & SNR & Tel./Sp.\\
 & [km~s$^{-1}$] & [km~s$^{-1}$] & [km~s$^{-1}$] & [km~s$^{-1}$] & [km~s$^{-1}$] & [km~s$^{-1}$] & [s]\\
\hline
6192.862570 & -65.99 & 0.22 & 0.12 & 56.23 & 0.10 & 0.07 & 900 & 52 & ESO/F\\
6193.875834 & -52.79 & 0.20 & 0.07 & 45.11 & 0.10 & -0.01 & 900 & 53 & ESO/F\\
6194.851951 & -39.05 & 0.25 & 0.11 & 33.69 & 0.08 & 0.002 & 750 & 42 & ESO/F\\  
6290.827518 & 24.44 & 0.18 & 0.16 & -19.20 & 0.08 & 0.007 & 800 & 45 & ESO/F\\
6291.741635 & 12.82 & 0.23 & -0.57 & -10.18 & 0.08 & -0.05 & 800 & 45& ESO/F\\   
6333.675179 & 27.29 & 0.21 & -0.08 & -21.95 & 0.14 & 0.09 & 1100 & 30 & CTIO/CH s\\
6338.539704 & 39.38 & 0.18 & 0.06 & -32.05 & 0.09 & -0.04 & 1100 & 30 & CTIO/CH s\\
6342.539638 & 28.41 & 0.61 & 0.03 & -23.51 & 0.22 & -0.12 & 750 & 50 & CTIO/CH f\\
6369.628830 & 19.67 & 0.61 & 0.33 & -15.86 & 0.17 & -0.007 & 750 & 50 & CTIO/CH f\\
6374.532803 & -73.75 & 0.64 & -0.32 & 61.70 & 0.23 & 0.18 & 750 & 50 & CTIO/CH f\\
6377.555037 & -42.79 & 0.44 & -0.07 & 35.81 & 0.19 & -0.09 & 750 & 50 & CTIO/CH f\\
6386.504040 & 29.69 & 0.52 & 0.34 & -24.22 & 0.23 & -0.02 & 750 & 50 & CTIO/CH f\\ 
6400.485080 & -73.67 & 0.84 & -0.96 & 60.09 & 0.22 & -0.84 & 750 & 50 & CTIO/CH f\\
6417.457549 & 39.29 & 0.61 & 0.15 & -32.29 & 0.23 & 0.06 & 750 & 50 & CTIO/CH f\\
6620.651060 & 25.64 & 0.16 & 0.15 & -20.25 & 0.10 & -0.05 & 1200 & 85 & ESO/C\\
6688.627369 & -72.95 & 0.47 & -0.18 & 61.12 & 0.13 & 0.15 & 750 & 50 & CTIO/CH f\\
6711.576390 & -21.82 & 0.35 & -0.02 & 18.51 & 0.15 & 0.06 & 750 & 50 & CTIO/CH f\\
6727.525042 & 34.14 & 0.17 & -0.06 & -27.38 & 0.12 & 0.09 & 1200 & 85 & EUL/C\\
6728.508227 & 36.64 & 0.19 & -0.08 & -29.53 & 0.10 & 0.02 & 1500 & 98 & EUL/C\\
6732.514401 & 38.00 & 0.16 & 0.09 & -30.57 & 0.12 & -0.02 & 1200 & 85 & EUL/C\\
6754.479720 & 36.17 & 0.32 & 0.25 & -29.51 & 0.19 & 0.16 & 750 & 50 & CTIO/CH f\\
6940.780201 & 38.96 & 0.30 & -0.22 & -31.59 & 0.20 & 0.03 & 1200 & 85 & EUL/C\\
6941.817413 & 37.19 & 0.17 & -0.08 & -30.06 & 0.12 & -0.04 & 1200 & 85 & EUL/C \\
\hline
\label{RV_table_051753}
\end{tabular}
\end{table}

\begin{table}
\caption{RV measurements for V643 Ori.}
\centering
\begin{tabular}{c c c c c c c c c c}
\hline \hline
BJD-2450000 & $RV_{1}$ & $\sigma_\mathrm{RV_{1}}$ & $O-C_{1}$ & $RV_{2}$ & $\sigma_\mathrm{RV_{2}}$ & $O-C_{2}$ & $T_\mathrm{exp}$ & SNR & Tel./Sp.\\
 & [km~s$^{-1}$] & [km~s$^{-1}$] & [km~s$^{-1}$] & [km~s$^{-1}$] & [km~s$^{-1}$] & [km~s$^{-1}$] & [s]\\
\hline 
4727.292831 & 63.68 & 0.33 & 0.00 & -32.44 & 0.93 & 0.00 & 900 & 45 & AAT/U\\
6178.865475 & 7.38 & 1.11 & 0.29 & 61.31 & 2.15 & -1.62 & 500 & 22 & EUL/C\\
6194.895866 & 63.06 & 0.39 & -0.001 & -32.45 & 0.49 & -0.29 & 360 & 67 & ESO/F\\ 
6196.855779 & 64.22 & 0.46 & -0.001 & -33.59 & 0.61 & 0.54 & 480 & 77 & ESO/F\\ 
6242.782295 & 53.44 & 0.83 & 0.06 & -16.91 & 1.76 & -1.12 & 300 & 34 & EUL/C\\
6271.757945 & -4.91 & 0.30 & -0.08 & 83.90 & 0.91 & 0.70 & 660 & 44 & CTIO/CH s\\
6291.808192 & 40.34 & 0.44 & -0.49 & 5.47 & 0.98 & -0.12 & 450 & 70 & ESO/F\\ 
6292.762692 & 44.55 & 0.36 & -0.10 & -0.92 & 0.60 & -0.04 & 600 & 80 & ESO/F\\
6295.789583 & 55.05 & 0.30 & -0.11 & -19.03 & 0.82 & -0.27 & 660 & 45 & CTIO/CH s\\
6325.630850 & -7.10 & 0.22 & 0.07 & 87.27 & 0.89 & 0.09 & 660 & 45 & CTIO/CH s\\
6346.580128 & 49.93 & 0.49 & -0.02 & -8.97 & 1.29 & 0.93 & 900 & 42 & EUL/C\\
6348.566506 & 56.01 & 0.51 & -0.18 & -19.39 & 1.96 & 1.10 & 300 & 32 & EUL/C\\
6350.655006 & 61.29 & 0.56 & 0.28 & -30.75 & 1.45 & -2.07 & 300 & 32 & EUL/C\\
6374.555008 & -0.36 & 0.42 & -0.60 & 74.70 & 1.06 & 0.15 & 450 & 60 & CTIO/CH f\\
6376.530285 & -4.97 & 0.39 & -0.41 & 82.42 & 1.26 & -0.29 & 450 & 62 & CTIO/CH f\\
6382.573157 & -7.61 & 0.57 & -0.21 & 87.90 & 1.40 & 0.35 & 450 & 60 & CTIO/CH f\\
6390.535449 & 15.60 & 0.36 & 0.63 & 51.02 & 1.33 & 1.51 & 450 & 65 & CTIO/CH f\\
6399.499802 & 51.72 & 0.29 & -0.08 & -13.01 & 1.20 & 0.07 & 450 & 64 & CTIO/CH f\\
6407.509660 & 63.91 & 0.53 & -0.41 & -34.25 & 1.40 & 0.09 & 450 & 62 & CTIO/CH f\\
6418.455726 & 36.45 & 0.66 & 2.71 & 17.53 & 0.72 & -0.06 & 450 & 60 & CTIO/CH f\\
6556.897775 & 52.31 & 0.50 & 0.18 & -12.61 & 1.65 & 1.01 & 600 & 40 & CTIO/CH s\\
6560.810923 & 61.75 & 0.45 & -0.09 & -30.18 & 1.03 & -0.07 & 600 & 40 & CTIO/CH s\\
6690.603649 & -3.78 & 0.27 & -0.12 & 82.49 & 1.40 & 1.31 & 450 & 60 & CTIO/CH f\\
6697.624454 & -6.81 & 0.27 & -0.09 & 85.88 & 0.56 & -0.51 & 450 & 63 & CTIO/CH f\\
6721.542967 & 64.22 & 0.26 & -0.25 & -35.08 & 0.68 & -0.48 & 450 & 62 & CTIO/CH f\\
6728.515786 & 51.71 & 0.27 & 0.44 & -12.51 & 1.23 & -0.32 & 450 & 64 & CTIO/CH f\\
6737.537820 & 13.98 & 0.25 & -0.10 & 53.34 & 0.89 & 2.34 & 450 & 65 & CTIO/CH f\\
6746.506415 & -8.33 & 0.24 & 0.02 & 87.71 & 0.78 & -1.44 & 450 & 60 & CTIO/CH f\\
6765.509722 & 48.49 & 0.49 & -0.15 & -6.58 & 1.27 & 1.13 & 450 & 60 & CTIO/CH f\\ 
\hline
\label{RV_table_V643}
\end{tabular}
\end{table}

\begin{table}
\caption{RV measurements for A-061016.}
\centering
\begin{tabular}{c c c c c c c c c c}
\hline \hline
BJD-2450000 & $RV_{1}$ & $\sigma_\mathrm{RV_{1}}$ & $O-C_{1}$ & $RV_{2}$ & $\sigma_\mathrm{RV_{2}}$ & $O-C_{2}$ & $T_\mathrm{exp}$ & SNR & Tel./Sp.\\
 & [km~s$^{-1}$] & [km~s$^{-1}$] & [km~s$^{-1}$] & [km~s$^{-1}$] & [km~s$^{-1}$] & [km~s$^{-1}$] & [s]\\
\hline
4043.826621 & 26.35 & 0.17 & -0.09 & 48.65 & 0.34 & -0.04 & 1200 & 55 & ESO/H\\
5846.822744 & 24.66 & 0.21 & -0.29 & 50.45 & 0.37 & 0.15 & 360 & 25 & EUL/C\\
5987.041087 & 61.46 & 0.19 & -0.16 & 13.88 & 0.45 & 0.12 & 420 & 35 & EUL/C\\
5988.036491 & 61.03 & 0.20 & -0.16 & 14.45 & 0.47 & 0.25 & 420 & 35 & EUL/C\\
5989.054502 & 60.60 & 0.20 & -0.11 & 14.88 & 0.42 & 0.22 & 500 & 40 & EUL/C\\
6081.540374 & 16.34 & 0.21 &  0.22 & 59.08 & 0.39 & -0.021 & 360 & 25 & EUL/C\\
6083.456183 & 16.27 & 0.27 &  0.16 & 59.00 & 0.55 & -0.11 & 360 & 25 & EUL/C\\
6160.908594 & 61.60 & 0.27 &  0.28 & 13.98 & 0.56 & -0.07 & 480 & 38 & EUL/C\\
6161.904393 & 62.13 & 0.47 &  0.33 & 13.45 & 0.77 & -0.12 & 480 & 38 & EUL/C\\
6178.887864 & 64.53 & 0.17 &  0.13 & 10.89 & 0.30 & -0.03 & 1200 & 55 & ESO/H\\
6180.824916 & 64.05 & 0.21 &  0.29 & 11.36 & 0.47 & -0.28 & 450 & 36 & EUL/C\\
6194.871465 & 57.98 & 0.21 &  0.22 & 17.44 & 0.48 & -0.05 & 750 & 70 & ESO/F\\
6195.896634 & 57.42 & 0.25 &  0.26 & 18.00 & 0.43 & -0.09 & 600 & 65 & ESO/F\\
6237.762240 & 29.91 & 0.21 &  0.29 & 45.00 & 0.41 & -0.65 & 600 & 40 & EUL/C\\
6241.858569 & 27.82 & 0.20 &  0.39 & 47.54 & 0.35 & -0.29 & 600 & 40 & EUL/C\\
6290.859929 & 16.80 & 0.29 &  0.02 & 58.54 & 0.44 & 0.21 & 500 & 55 & ESO/F\\
6291.782923 & 16.93 & 0.26 &  0.02 & 58.45 & 0.45 & 0.26 & 500 & 60 & ESO/F\\
6291.790762 & 16.90 & 0.25 &  -0.01 & 58.43 & 0.43 & 0.23 & 500 & 55 & ESO/F\\
6292.745356 & 16.96 & 0.27 &  -0.11 & 58.23 & 0.38 & 0.19 & 500 & 57 & ESO/F\\
6296.732647 & 17.65 & 0.22 &  0.02 & 57.47 & 0.44 & 0.01 & 900 & 40 & CTIO/CH s\\ 
6304.719226 & 20.17 & 0.22 &  0.01 & 54.91 & 0.44 & -0.03 & 900 & 45 & CTIO/CH s\\  
6312.612448 & 23.86 & 0.26 & -0.02 & 51.33 & 0.43 & 0.09 & 900 & 43 & CTIO/CH s\\
6345.573572 & 50.13 & 0.17 &  -0.10 & 25.42 & 0.54 & 0.32 & 600 & 40 & EUL/C \\
6347.667121 & 51.91 & 0.19 &  -0.12 & 23.54 & 0.47 & 0.23 & 600 & 40 & EUL/C\\
6349.657388 & 53.49 & 0.22 & -0.20 & 21.77 & 0.43 & 0.12 & 600 & 40 & EUL/C\\
6398.470455 & 55.31 & 0.21 & -0.15 & 19.95 & 0.42 & 0.07 & 900 & 56 & EUL/C\\
6497.944224 & 17.87 & 0.23 & -0.19 & 57.26 & 0.28 & 0.08 & 600 & 40 & EUL/C\\
6518.878550 & 27.81 & 0.27 & -0.12 & 47.31 & 0.29 & 0.09 & 420 & 45 & ESO/F\\
6519.899841 & 28.50 & 0.25 & -0.12 & 46.43 & 0.26 & -0.09 & 480 & 48 & ESO/F\\
6520.918986 & 29.19 & 0.23 & -0.22 & 45.57 & 0.24 & -0.24 & 480 & 50 & ESO/F\\
6953.734346 & 56.59 & 0.22 & -0.04 & 18.45 & 0.39 & -0.14 & 900 & 50 & CTIO/CH s\\
7005.756787 & 50.30 & 0.20 & -0.29 & 24.58 & 0.44 & -0.03 & 900 & 52 & CTIO/CH s\\
7049.680142 & 23.30 & 0.20 & -0.33 & 51.62 & 0.46 & 0.14 & 900 & 48 & CTIO/CH s\\ 
\hline
\label{RV_table_061016}
\end{tabular}
\end{table}

\begin{table}
\caption{RV measurements for A-062926.}
\centering
\begin{tabular}{c c c c c c c c c c}
\hline \hline
BJD-2450000 & $RV_{1}$ & $\sigma_\mathrm{RV_{1}}$ & $O-C_{1}$ & $RV_{2}$ & $\sigma_\mathrm{RV_{2}}$ & $O-C_{2}$ & $T_\mathrm{exp}$ & SNR & Tel./Sp.\\
 & [km~s$^{-1}$] & [km~s$^{-1}$] & [km~s$^{-1}$] & [km~s$^{-1}$] & [km~s$^{-1}$] & [km~s$^{-1}$] & [s]\\
\hline
5962.537810 & -38.66 & 0.09 & 0.05 & 53.08 & 0.14 & 0.02 & 1600 & 55 & ESO/F\\
6290.809923 & 35.34 & 0.24 & 0.13 & -20.98 & 0.20 & -0.11 & 800 & 32 & ESO/F\\
6291.732238 & 28.45 & 0.11 & 0.01 & -14.07 & 0.12 & 0.03 & 800 & 35 & ESO/F\\
6292.673654 & 22.21 & 0.18 & 0.12 & -7.56 & 0.21 & 0.19 & 800 & 32 & ESO/F\\
6342.6064767 & 43.09 & 0.19 & 0.39 & -29.04 & 0.19 & -0.17 & 900 & 32 & CTIO/CH f\\
6366.605824 & 65.05 & 0.31 & -0.28 & -51.22 & 0.54 & 0.01 & 900 & 32 & CTIO/CH f\\
6370.563467 & 30.53 & 0.25 & -0.09 & -16.39 & 0.25 & 0.14 & 900 & 32 & CTIO/CH f\\
6381.543453 & -27.36 & 0.31 & -0.03 & 41.15 & 0.29 & -0.26 & 900 & 32 & CTIO/CH f\\
6381.586528 & -27.42 & 0.06 & 0.07 & 41.65 & 0.34 & -0.17 & 780 & 30 & ESO/F\\
6382.541565 & -31.88 & 0.09 & -0.43 & 45.58 & 0.29 & -0.22 & 780 & 30 & ESO/F\\
6383.547749 & -35.18 & 0.13 & 0.14 & 49.62 & 0.21 & -0.05 & 780 & 31 & ESO/F\\
6397.549626 & 26.27 & 0.21 & -0.08 & -12.52 & 0.35 & -0.25 & 900 & 32 & CTIO/CH f\\ 
6412.515243 & -40.29 & 0.16 & -0.001 & 54.52 & 0.31 & -0.15 & 900 & 32 & CTIO/CH f\\
6423.482839 & 29.38 & 0.27 & -0.09 & -15.17 & 0.25 & 0.21 & 900 & 32 & CTIO/CH f\\
6423.496239 & 29.45 & 0.27 & 0.07 & -15.26 & 0.25 & 0.03 & 900 & 32 & CTIO/CH f\\
6429.490837 & -4.82 & 0.40 & 0.39 & 19.46 & 0.55 & -0.09 & 900 & 32 & ESO/F\\
6698.634607 & -29.56 & 0.19 & 0.01 & 43.68 & 0.19 & 0.02 & 900 & 32 & CTIO/CH f\\
6711.591811 & 45.80 & 0.15 & -0.04 & -31.95 & 0.21 & -0.20 & 900 & 32 & CTIO/CH f\\
6733.532766 & 70.81 & 0.14 & 0.005 & -56.74 & 0.22 & -0.03 & 900 & 32 & CTIO/CH f\\
6753.568213 & -37.57 & 0.14 & 0.12 & 51.92 & 0.31 & 0.15 & 900 & 32 & CTIO/CH f\\
\hline
\label{RV_table_062926}
\end{tabular}
\end{table}

\begin{table}
\caption{RV measurements for A-065114.}
\centering
\begin{tabular}{c c c c c c c c c c}
\hline \hline
BJD-2450000 & $RV_{1}$ & $\sigma_\mathrm{RV_{1}}$ & $O-C_{1}$ & $RV_{2}$ & $\sigma_\mathrm{RV_{2}}$ & $O-C_{2}$ & $T_\mathrm{exp}$ & SNR & Tel./Sp.\\
 & [km~s$^{-1}$] & [km~s$^{-1}$] & [km~s$^{-1}$] & [km~s$^{-1}$] & [km~s$^{-1}$] & [km~s$^{-1}$] & [s]\\
\hline
6245.870153 & -14.51 & 0.53 & 0.09 & 57.63 & 0.94 & -1.88 & 600 & 37 & CTIO/CH s\\
6268.768114 & 62.86 & 0.68 & -0.22 & -19.84 & 0.63 & -0.08 & 600 & 32 & CTIO/CH s\\
6290.822314 & -19.69 & 0.46 & 0.19 & 64.85 & 0.54 & -0.06 & 400 & 74 & ESO/F\\
6291.770275 & -22.60 & 0.69 & -0.25 & 67.84 & 0.41 & 0.41 & 400 & 75 & ESO/F\\
6292.726561 & -24.15 & 0.59 & -0.16 & 69.77 & 0.74 & 0.66 & 400 & 55 & ESO/F\\
6338.633898 & -24.69 & 0.63 & -0.48 & 69.05 & 0.75 & -0.28 & 450 & 68 & CTIO/CH f\\ 
6344.584194 & -2.29 & 0.72 & 0.18 & 47.02 & 0.58 & -0.11 & 450 & 66 & CTIO/CH f\\
6366.620900 & 44.41 & 0.67 & -0.53 & -1.29 & 0.60 & -0.04 & 450 & 70 & CTIO/CH f\\
6384.519660 & -18.84 & 0.59 & 0.11 & 63.68 & 0.89 & -0.27 & 450 & 70 & CTIO/CH f\\
6397.477012 & 55.95 & 0.57 & 0.01 & -12.24 & 0.64 & 0.23 & 480 & 42 & EUL/C\\
6398.479098 & 60.29 & 0.41 & 0.03 & -16.33 & 0.76 & 0.56 & 480 & 44 & EUL/C\\
6573.873445 & 65.49 & 0.53 & 0.51 & -20.93 & 1.78 & 0.75 & 600 & 25 & CTIO/CH s\\
6580.854637 & 60.91 & 0.62 & -0.09 & -18.44 & 0.66 & -0.81 & 600 & 28 & CTIO/CH s\\
6689.659346 & -16.31 & 0.57 & 0.289 & 59.58 & 0.86 & -1.97 & 450 & 68 & CTIO/CH f\\
6707.570441 & 68.81 & 0.44 & -0.12 & -26.35 & 0.50 & -0.63 & 450 & 68 & CTIO/CH f\\
6723.526522 & -11.63 & 0.75 & -0.70 & 57.26 & 0.67 & 1.48 & 450 & 66 & CTIO/CH f\\
6732.547564 & -18.57 & 0.63 & 0.21 & 62.73 & 0.82 & -1.05 & 450 & 67 & CTIO/CH f\\
6742.547056 & 39.21 & 0.58 & -0.56 & 5.60 & 0.73 & 1.58 & 450 & 70 & CTIO/CH f\\
6743.543691 & 45.96 & 0.44 & 0.14 & -0.72 & 0.07 & 1.43 & 450 & 70 & CTIO/CH f\\
\hline
\label{RV_table_065114}
\end{tabular}
\end{table}

\begin{table}
\caption{RV measurements for A-090232.}
\centering
\begin{tabular}{c c c c c c c c c c}
\hline \hline
BJD-2450000 & $RV_{1}$ & $\sigma_\mathrm{RV_{1}}$ & $O-C_{1}$ & $RV_{2}$ & $\sigma_\mathrm{RV_{2}}$ & $O-C_{2}$ & $T_\mathrm{exp}$ & SNR & Tel./Sp.\\
 & [km~s$^{-1}$] & [km~s$^{-1}$] & [km~s$^{-1}$] & [km~s$^{-1}$] & [km~s$^{-1}$] & [km~s$^{-1}$] & [s]\\
\hline
6695.793414 & -21.36 & 0.63 & 0.89 & 70.11 & 1.24 & -0.68 & 750 & 55 & CTIO/CH f\\
6715.664167 & -13.04 & 0.52 & -0.52 & 60.69 & 0.97 & 0.11 & 750 & 52 & CTIO/CH f\\
6720.638695 & -19.75 & 0.53 & -1.08 & 66.15 & 1.69 & -0.88 & 750 & 57 & CTIO/CH f\\
6728.524254 & 75.85 & 0.62 & -0.51 & -32.69 & 1.03 & -0.21 & 750 & 54 & CTIO/CH f\\
6743.606859 & 10.41 & 0.63 & 0.73 & 37.06 & 1.52 & -0.27 & 750 & 55 & CTIO/CH f\\
6749.545619 & 76.25 & 0.62 & -0.32 & -32.91 & 0.95 & -0.21 & 750 & 53 & CTIO/CH f\\
6757.593831 & -15.80 & 0.45 & -0.03 & 64.10 & 0.99 & 0.11 & 750 & 58 & CTIO/CH f\\
6770.472093 & 77.13 & 0.51 & 0.52 & -32.57 & 0.95 & 0.17 & 750 & 55 & CTIO/CH f\\
6778.502956 & -15.49 & 0.59 & 1.21 & 65.56 & 1.14 & 0.59 & 750 & 53 & CTIO/CH f\\
6797.477627 & 6.98 & 0.74 & -1.05 & 39.94 & 1.76 & 0.88 & 750 & 56 & CTIO/CH f\\
6814.504378 & 63.60 & 0.53 & 0.06 & -18.79 & 1.19 & 0.27 & 750 & 52 & CTIO/CH f\\
\hline
\label{RV_table_090232}
\end{tabular}
\end{table}

\begin{table}
\caption{RV measurements for A-110814.}
\centering
\begin{tabular}{c c c c c c c c c c}
\hline \hline
BJD-2450000 & $RV_{1}$ & $\sigma_\mathrm{RV_{1}}$ & $O-C_{1}$ & $RV_{2}$ & $\sigma_\mathrm{RV_{2}}$ & $O-C_{2}$ & $T_\mathrm{exp}$ & SNR & Tel./Sp.\\
 & [km~s$^{-1}$] & [km~s$^{-1}$] & [km~s$^{-1}$] & [km~s$^{-1}$] & [km~s$^{-1}$] & [km~s$^{-1}$] & [s]\\
\hline
5961.664023 & -27.551 & 0.245 & -0.082 & 46.015 & 0.146 & -0.03 & 900 & 65 & ESO/F\\
6080.500138 &  51.434 & 0.226 & -0.039 & -30.182 & 0.221 & 0.32 & 660 & 32 & EUL/C\\
6082.505315 &  50.626 & 0.297 & 0.023 & -29.252 & 0.197 & 0.41 & 660 & 34 & EUL/C\\
6291.812522 & 26.060 & 0.261 & -0.170 & -5.950 & 0.121 & 0.03 & 800 & 58 & ESO/F\\
6292.709716 & 29.210 & 0.281 & 0.135 & -8.755 & 0.100 & -0.02 & 800 & 57 & ESO/F\\
6292.766433 & 29.42 & 0.28 & 0.17 & -8.89 & 0.14 & 0.01 & 800 & 55 & ESO/F\\
6335.739335 & -24.04 & 0.14 & -0.12 & 42.36 & 0.11 & -0.17 & 1200 & 40 & CTIO/CH s\\
6342.763926 & -32.24 & 0.23 & -0.17 & 49.98 & 0.14 & -0.06 & 750 & 54 & CTIO/CH f\\ 
6344.611290 & -31.74 & 0.17 & -0.01 & 50.16 & 0.12 & 0.07 & 1200 & 40 & CTIO/CH s\\
6348.726915 & -27.49 & 0.39 & 0.38 & 46.31 & 0.19 & -0.05 & 780 & 35 & EUL/C\\
6350.793830 & -23.83 & 0.28 & 0.41 & 42.69 & 0.16 & -0.15 & 780 & 35 & EUL/C\\
6376.771248 & 49.06 & 0.35 & 0.19 & -28.08 & 0.14 & 0.29 & 750 & 54 & CTIO/CH f\\  
6393.807280 & 29.08 & 0.29 & -0.01 & -9.29 & 0.19 & -0.07 & 750 & 54 & CTIO/CH f\\ 
6424.666037 & -26.85 & 0.34 & 0.04 & 45.04 & 0.12 & 0.03 & 750 & 54 & CTIO/CH f\\ 
6721.571753 & -31.39 & 0.28 & 0.07 & 49.55 & 0.12 & 0.11 & 750 & 54 & CTIO/CH f\\
6728.537163 & -20.50 & 0.28 & 0.21 & 39.08 & 0.09 & 0.05 & 750 & 54 & CTIO/CH f\\ 
6747.542720 & 38.96 & 0.17 & 0.03 & -18.58 & 0.17 & 0.16 & 750 & 54 & CTIO/CH f\\ 
6752.545109 & 48.53 & 0.23 & 0.02 &  -28.19 & 0.13 & -0.16 & 750 & 54 & CTIO/CH f\\ 
6770.485699 & 26.93 & 0.25 &  -0.24 & -7.37 & 0.09 & -0.01 & 750 & 54 & CTIO/CH f\\ 
6789.486760 & -28.32 & 0.26 & 0.07 & 46.55 & 0.13 & 0.08 & 750 & 54 & CTIO/CH f\\ 
6832.512188 & 51.28 & 0.21 & -0.04 & -31.03 & 0.11 & -0.28 & 750 & 54 & CTIO/CH f\\ 
\hline
\label{RV_table_110814}
\end{tabular}
\end{table}

\twocolumn

\label{lastpage}
\end{document}